\documentclass[useAMS,a4paper,fleqn,usenatbib]{mn2e}
\usepackage{times}
\usepackage{amsmath}
\usepackage{amssymb}
\usepackage{amsbsy}
\usepackage{bmpsize}
\usepackage{graphicx}
\usepackage{subfigure}
\usepackage{paralist}
\usepackage{mathrsfs}
\usepackage{booktabs}
\usepackage{tabularx}
\usepackage{color}
\definecolor{darkred}{rgb}{0.55, 0.0, 0.0}
\definecolor{darkblue}{rgb}{0.0, 0.0, 0.55}
\usepackage[colorlinks=true,linkcolor=darkred,citecolor=darkblue]{hyperref}

\newcommand{\specialcell}[2][t]{%
	\bgroup
	\def\arraystretch{1.2}
  \begin{tabular}[#1]{@{}l@{}}#2\end{tabular}\egroup}

\let\boldgrk=\gkvecten
\let\boldgrksc=\gkvecseven

\def\gkthing#1{{\mathchoice%
	{\hbox{{\boldgrk\char#1}}}
	{\hbox{{\boldgrk\char#1}}}
	{\hbox{{\boldgrksc\char#1}}}
	{\hbox{{\boldgrksc\char#1}}}}}

\def\vtheta{\gkthing{18}}

\def\vOmega{{\bf\Omega}}
\def\vDelta{{\bf\Delta}}
{\newif\ifnotend
\notendtrue
\def\veclist{ABCDEFGHIJKLMNOPQRSTUVWXYZabcdefghijklmnopqrstuvwxyz.}
\def\top#1#2.{#1}
\def\tail#1#2.{#2.}
\loop\expandafter\xdef\csname v\expandafter\top\veclist\endcsname%
{{\noexpand\bf\expandafter\top\veclist}}
\edef\veclist{\expandafter\tail\veclist}
\if\veclist.\notendfalse\fi\ifnotend\repeat}

\def\kpc{\,{\rm kpc}}

\def\kms{\,{\rm km\,s^{-1}}}
\def\rad{\,\rm rad}
\def\Gyr{\,{\rm Gyr}}

\def\percent{\text{ per cent}}
\def\degrees{\text{ degrees}}
\def\d{{\rm d}}
\def\half{{\textstyle{\frac12}}}
\def\sqrthalf{{\textstyle{\sqrt{\frac12}}}}
\newcommand{\mat}[1]{\mathbfss{#1}}
\newcommand{\bs}[1]{\boldsymbol{#1}}
\def\df{{\sc df}}
\title[Action estimation in triaxial potentials]{A fast algorithm for estimating actions in triaxial potentials}
\author[J. L. Sanders \& J. Binney]{Jason L. Sanders$^{1,2}$\thanks{E-mail: jls@ast.cam.ac.uk} \& James Binney$^1$\\
$^1$Rudolf Peierls Centre for Theoretical Physics, Keble Road, Oxford, OX1 3NP, UK\\
$^2$Institute of Astronomy, Madingley Road, Cambridge, CB3 0HA}

\pagerange{\pageref{firstpage}--\pageref{lastpage}} \pubyear{2014}

\begin{document}
\maketitle
\begin{abstract}
We present an approach to approximating rapidly the actions in a general
triaxial potential.  The method is an extension of the axisymmetric approach
presented by \cite{Binney2012}, and operates by assuming that the true
potential is locally sufficiently close to some St\"ackel potential. The
choice of St\"ackel potential and associated ellipsoidal coordinates is
tailored to each individual input phase-space point. We investigate the
accuracy of the method when computing actions in a triaxial
Navarro-Frenk-White potential. The speed of the algorithm comes at the
expense of large errors in the actions, particularly for the box orbits.
However, we show that the method can be used to recover the observables of
triaxial systems from given distribution functions to sufficient accuracy for
the Jeans equations to be satisfied. Consequently, such models could be used
to build models of external galaxies as well as triaxial components of our
own Galaxy. When more accurate actions are required, this procedure can be
combined with torus mapping to produce a fast convergent scheme for
action estimation.
\end{abstract}
\label{firstpage}
\section{Introduction}

The haloes that form in baryon-free cosmological simulations almost always
have triaxial shapes \citep{JingSuto2002,Allgood2006,VeraCiro2011}. When baryons are added to the simulations, many dark haloes become more spherical \citep{Kazantzidis2004,Bailin2005,Valluri2010}, but the most successful current models suggest our Galaxy's dark halo is triaxial \citep{LawMajewski2010,VeraCiro2013}. Moreover, there is considerable observational evidence that the so-called ``cored'', slowly-rotating elliptical galaxies are generically triaxial \citep{Atlas3D}.
Hence dynamical models of triaxial stellar systems are of considerable astronomical interest.

The first triaxial models were made by violent relaxation of an $N$-body model
\citep{AarsethBinney}, and these models prompted \cite{Schwarzschild1979} to
develop the technique of orbit superposition so triaxial models with
prescribed density profiles could be constructed.  Schwarzschild's work gave significant insight into how triaxial systems work for
the first time, and this insight was enhanced by \cite{deZeeuw1985}, who showed that St\"ackel potentials provided analytic models of orbits in a very interesting class of triaxial systems. The most important subsequent development in the study of triaxial systems was the demonstration by \cite{MerrittValluri} that when a triaxial system lacks a homogeneous core, as real galaxies do, box orbits tend to become centrophobic resonant box orbits.

Work on axisymmetric models in the context of our Galaxy has increased
awareness of the value in stellar dynamics of the intimately related concepts
of the Jeans' theorem and action integrals. \cite{Delhaye} established that
the space of quasi-periodic orbits in galactic potentials is
three-dimensional. Jeans' theorem tells us that any non-negative function $f$
on this space provides an equilibrium stellar system. The key to gaining
access to the observable properties of this tantalising array of stellar
systems, is finding a practical coordinate system for orbit space. A
coordinate system for orbit space comprises a set of three functions
$I_i(\vx,\vv)$ that are constant along any orbit in the gravitational
potential $\Phi(\vx)$ of the equilibrium system. A major difficulty is that
in the case of a self-consistent system
$\Phi(\vx)$ has to be determined from $f(\vI)$ by computing the model's
density, and the latter can be computed only when $\Phi(\vx)$ is known. Hence
the computation of $\Phi(\vx)$ has to be done iteratively, and expressions
are needed for the $I_i$ that are valid in {\it any} reasonable potential
$\Phi(\vx)$, not merely the potential of the equilibrium model.

If the system is axisymmetric, the energy $E$ and component of angular
momentum $L_z$ are integrals that are defined for any axisymmetric potential
$\Phi(R,z)$, and equilibrium models of axisymmetric systems have been
constructed from distribution functions (\df{}s) of the form $f(E,L_z)$
\citep{PrendergastTomer,Wilson1975,Rowley1988}.  However, these two-integral
models are not generic, and they are much harder to construct than generic
models when the \df{} is specified as a function $f(\vJ)$ of the actions
\citep{Binney2014}. Moreover, knowledge of the \df{} as a function of the
actions is the key to Hamiltonian perturbation theory, and the ability to
perturb models is crucial if we are to really understand how galaxies work,
and evolve over time. Actions are also the key to modelling stellar streams,
which are themselves promising probes of our Galaxy's distribution of dark
matter \citep{Tremaine1999, HelmiWhite1999, EyreBinney2011, SandersBinney2013}.

Action integrals constitute uniquely advantageous coordinates for orbit
space, which is often called action space because its
natural Cartesian coordinates are the actions. Actions can be defined for any
quasi-periodic orbit and, uniquely among isolating integrals, they can be
complemented by canonically conjugate coordinates, the angle variables
$\theta_i$. These have the convenient properties of (i) increasing linearly
in time, so
 \[
\theta_i(t)=\theta_i(0)+\Omega_i(\vJ)t,
\]
and (ii) being periodic such that any ordinary phase-space coordinate such as
$x$ satisfies $x(\vtheta,\vJ)=x(\vtheta+2\pi\vm,\vJ)$, where $\vm$ is any
triple of integers.

The discussion above amply motivates the quest for algorithms that yield
angle-action coordinates $(\vtheta,\vJ)$ given ordinary phase-space
coordinates $(\vx,\vv)$, and vice versa. These algorithms are usefully
divided into convergent and non-convergent algorithms. Convergent algorithms
yield approximations to the desired quantity that can achieve any desired
accuracy given sufficient computational resource, whereas non-convergent
algorithms provide, more cheaply, an approximation of uncontrolled accuracy.
Torus mapping \citep{KaasalainenBinney,McMillanBinney2008} is a convergent
algorithm that yields $\vx(\vtheta,\vJ)$ and $\vv(\vtheta,\vJ)$, while
\cite{SandersBinney2014} introduced a convergent algorithm for
$\vtheta(\vx,\vv)$ and $\vJ(\vx,\vv)$. Both algorithms work by constructing
the generating function for the canonical mapping of some ``toy'' analytic
system of angle-action variables into the real phase space. Torus mapping has
been demonstrated only in two dimensions, but both axisymmetric and two-dimensional static barred
potentials have been successfully handled, and there is no evident obstacle
to generalising to the three-dimensional case. \cite{SandersBinney2014}
treated the triaxial case, but the restriction to lower dimensions and
axisymmetry is trivial.

These convergent algorithms are numerically costly, and, in the axisymmetric
case, non-convergent algorithms have been used extensively, especially for
extracting observables from a \df{} $f(\vJ)$. These extractions require
$\vJ$ to be evaluated at very many phase-space points, and speed is more
important than accuracy. The adiabatic approximation \citep{Binney2010} has
been extensively used in modelling the solar neighbourhood
\citep[e.g.][]{SchoenrichBinney2009} but its validity is restricted to orbits
that keep close to the Galactic plane. St\"ackel fitting \citep{Sanders2012}
has been successfully used to model stellar streams \citep{Sanders2014}. This method estimates the actions as those in the best-fitting St\"ackel potential for the local region a give orbit probes. \cite{Sanders2015}
shows that is is less cost-effective than the ``St\"ackel fudge'' that was introduced by
\cite{Binney2012}. \cite{Binney2012b} used the ``St\"ackel fudge'' to model the solar neighbourhood and to explore the first family of self-consistent stellar systems with specified $f(\vJ)$ \citep{Binney2014}. The St\"ackel fudge was recently used by
\cite{Piffl2014} to place by far the strongest available constraints on the
Galaxy's dark halo. In this paper we extend the St\"ackel fudge to triaxial
systems.

We begin in Section~\ref{Sec::StackPot} by showing how to find the actions in a triaxial St\"ackel potential. In Section~\ref{Sec::Triaxapproximation} we extend the St\"ackel fudge to
general triaxial potentials. In Section~\ref{Sec::Tests} we apply this
algorithm to a series of orbits in a triaxial Navarro-Frenk-White (NFW) potential, and in
Section~\ref{Sec::DF} we construct the first triaxial stellar systems with
specified \df{}s $f(\vJ)$, and demonstrate that, notwithstanding the
uncontrolled nature of the fudge as an approximation, the models satisfy the
Jeans equations to good accuracy. In Section~\ref{sec::twoup} we describe a
new convergent algorithm for obtaining $(\vJ,\btheta)$ from $(\vx,\vv)$.
Finally we conclude in Section~\ref{Sec::TriaxConclusions}.

\section{Triaxial St\"ackel potentials}\label{Sec::StackPot}
In this section, we show how actions can be found in a triaxial St\"ackel potential. The presentation here follows that given by \cite{deZeeuw1985}.

\subsection{Ellipsoidal coordinates}
Triaxial St\"ackel potentials are expressed in terms of ellipsoidal coordinates $(\lambda,\mu,\nu)$. These coordinates are related to the Cartesian coordinates $(x,y,z)$ as the three roots of the cubic in $\tau$
\begin{equation}
\frac{x^2}{(\tau+\alpha)}+\frac{y^2}{(\tau+\beta)}+\frac{z^2}{(\tau+\gamma)} = 1,
\end{equation}
where $\alpha$, $\beta$ and $\gamma$ are constants defining the coordinate system. For the potential explored later, we choose to set $x$ as the major axis, $y$ as the intermediate axis and $z$ as the minor axis, such that $-\gamma\leq\nu\leq-\beta\leq\mu\leq-\alpha\leq\lambda$. Surfaces of constant $\lambda$ are ellipsoids, surfaces of constant $\mu$ are hyperboloids of one sheet (flared tubes of elliptical cross section that surround the $x$ axis), and surfaces of constant $\nu$ are hyperboloids of two sheets that have their extremal point on the $z$ axis. In the plane $z=0$, lines of constant $\lambda$ are ellipses with foci at $y=\pm\Delta_1\equiv\pm\sqrt{\beta-\alpha}$, whilst, in the plane $x=0$, lines of constant $\mu$ are ellipses with foci at $z=\pm\Delta_2\equiv\pm\sqrt{\gamma-\beta}$. The expressions for the Cartesian coordinates as a function of the ellipsoidal coordinates are
\begin{equation}
\begin{split}
x^2 &= \frac{(\lambda+\alpha)(\mu+\alpha)(\nu+\alpha)}{(\alpha-\beta)(\alpha-\gamma)},\\
y^2 &= \frac{(\lambda+\beta)(\mu+\beta)(\nu+\beta)}{(\beta-\alpha)(\beta-\gamma)},\\
z^2 &= \frac{(\lambda+\gamma)(\mu+\gamma)(\nu+\gamma)}{(\gamma-\beta)(\gamma-\alpha)}.
\end{split}
\label{Eq::CartFromTau}
\end{equation}
Note that a Cartesian coordinate $(x,y,z)$ gives a unique $(\lambda,\mu,\nu)$, whilst the point $(\lambda,\mu,\nu)$ corresponds to eight points in $(x,y,z)$. Therefore, we will only consider potentials with this symmetry i.e. triaxial potentials with axes aligned with the Cartesian axes.

The generating function, $S$, to take us between Cartesian, $(x,y,z,p_x,p_y,p_z)$, and ellipsoidal coordinates, $(\lambda,\mu,\nu,p_\lambda,p_\mu,p_\nu)$, is
\begin{equation}
S(p_x,p_y,p_z,\lambda,\mu,\nu) = p_x x(\lambda,\mu,\nu)+p_y y(\lambda,\mu,\nu)+p_z z(\lambda,\mu,\nu).
\end{equation}
Using $p_\tau = \upartial S/\upartial \tau$ we find, for instance,
\begin{equation}
\begin{split}
p_\lambda = &\frac{p_x}{2}\sqrt{\frac{(\mu+\alpha)(\nu+\alpha)}{(\alpha-\beta)(\alpha-\gamma)(\lambda+\alpha)}}\\+&\frac{p_y}{2}\sqrt{\frac{(\mu+\beta)(\nu+\beta)}{(\beta-\alpha)(\beta-\gamma)(\lambda+\beta)}}\\+&\frac{p_z}{2}\sqrt{\frac{(\mu+\gamma)(\nu+\gamma)}{(\gamma-\alpha)(\gamma-\beta)(\lambda+\gamma)}}.
\end{split}
\label{Eq::Example_ptau}
\end{equation}
There are similar equations for $p_\mu$ and $p_\nu$. Inversion of these three equations gives us expressions for $p_x,p_y$ and $p_z$ as functions of $p_\tau$ and $\tau$. For a general triaxial potential, $\Phi$, we can express the Hamiltonian, $H$, in terms of the ellipsoidal coordinates as
\begin{equation}
\begin{split}
H &= \half(p_x^2+p_y^2+p_z^2)+\Phi(x,y,z),\\
&=\half\Big(\frac{p_\lambda^2}{P_\lambda^2}+\frac{p_\mu^2}{P_\mu^2}+\frac{p_\nu^2}{P_\nu^2}\Big)+\Phi(\lambda,\mu,\nu).
\end{split}
\label{Eq::Hamiltonian}
\end{equation}
where
\begin{equation}
\begin{split}
P^2_\lambda &= \frac{(\lambda-\mu)(\lambda-\nu)}{4(\lambda+\alpha)(\lambda+\beta)(\lambda+\gamma)},\\
P^2_\mu &= \frac{(\mu-\nu)(\mu-\lambda)}{4(\mu+\alpha)(\mu+\beta)(\mu+\gamma)},\\
P^2_\nu &= \frac{(\nu-\mu)(\nu-\lambda)}{4(\nu+\alpha)(\nu+\beta)(\nu+\gamma)}.
\end{split}
\end{equation}

\subsection{St\"ackel potentials}
The most general triaxial St\"ackel potential, $\Phi_S$, can be written as
\begin{equation}
\Phi_S(\lambda,\mu,\nu) = \frac{f(\lambda)}{(\lambda-\mu)(\nu-\lambda)}+\frac{f(\mu)}{(\mu-\nu)(\lambda-\mu)}+\frac{f(\nu)}{(\nu-\lambda)(\mu-\nu)}.
\end{equation}
$\Phi_S$ is composed of three functions of one variable. Here we denote the three functions with the same letter, $f$, as their domains are distinct. Additionally, $f(\tau)$ must be differentiable everywhere and continuous at $\tau=-\alpha$ and $\tau=-\beta$ for $\Phi_S$ to be finite at $\lambda=\mu=-\alpha$ and $\mu=\nu=-\beta$. With this form for the potential we can solve the Hamilton-Jacobi equation \citep{deZeeuw1985}. We write $p_\tau=\upartial W/\upartial \tau$ and equate the Hamiltonian to the total energy, $E$, in equation~\eqref{Eq::Hamiltonian}. We then multiply through by $(\lambda-\mu)(\mu-\nu)(\nu-\lambda)$ to find
\begin{equation}
\begin{split}
&(\nu-\mu)\Big(2(\lambda+\alpha)(\lambda+\beta)(\lambda+\gamma)\Big(\frac{\upartial W}{\upartial \lambda}\Big)^2-f(\lambda)-\lambda^2E\Big)\\
+&(\lambda-\nu)\Big(2(\mu+\alpha)(\mu+\beta)(\mu+\gamma)\Big(\frac{\upartial W}{\upartial \mu}\Big)^2-f(\mu)-\mu^2E\Big)\\
+&(\mu-\lambda)\Big(2(\nu+\alpha)(\nu+\beta)(\nu+\gamma)\Big(\frac{\upartial W}{\upartial \nu}\Big)^2-f(\nu)-\nu^2E\Big)\\ &= 0.
\end{split}
\end{equation}
We make the Ansatz $W = \sum_\tau W_\tau(\tau)$ and define
\begin{equation}
U(\tau) = 2(\tau+\alpha)(\tau+\beta)(\tau+\gamma)\Big(\frac{\upartial W}{\upartial \tau}\Big)^2-f(\tau)-\tau^2E,
\end{equation}
such that the Hamilton-Jacobi equation becomes
\begin{equation}
(\nu-\mu)U(\lambda)
+(\lambda-\nu)U(\mu)
+(\mu-\lambda)U(\nu)= 0.
\end{equation}
Taking the second derivative of this expression with respect to $\tau = \{\lambda,\mu,\nu\}$ we find that
\begin{equation}
U(\tau)=a\tau-b,
\end{equation}
where $a$ and $b$ are constants. Therefore, the equations for the momenta can be written as
\begin{equation}
2(\tau+\alpha)(\tau+\beta)(\tau+\gamma)p_\tau^2=\tau^2 E -\tau a+b + f(\tau).
\label{Eq::EqnOfMotion}
\end{equation}
For an initial phase-space point, $(\vx_0,\vv_0)$, we find $\tau_0(\vx_0,\vv_0)$ and $p_{\tau 0}(\vx_0,\vv_0)$ using the coordinate transformations and can then find the integrals $a$ and $b$ by solving equation~\eqref{Eq::EqnOfMotion} \citep[see][for more details]{deZeeuw1985}. These integrals are related to the classical integrals $I_2$ and $I_3$ in a simple way. As $p_\tau$ is only a function of $\tau$,
the actions are then given by the 1D integrals
\begin{equation}
J_\tau = \frac{2}{\upi}\int_{\tau_-}^{\tau_+} \mathrm{d}\tau\,|p_\tau(\tau)|.
\label{Eq::action}
\end{equation}
where ($\tau_-,\tau_+$) are the roots of $p_\tau(\tau)=0$, which we find by
using Brent's method to find points where the right side of equation~\eqref{Eq::EqnOfMotion} vanishes. Note that for loop orbits we must divide the `radial' action by two ($J_\lambda$ for the short-axis loops and outer long-axis loops, $J_\mu$ for the inner long-axis loops). In Table~\ref{ActionMeaning}, we give the limits $(\tau_-,\tau_+)$ of the action integrals and the physical meaning of each of the actions for each of the four orbit classes.

\begin{table}
\caption{Actions in a triaxial St\"ackel potential. We give the limits of the action integrals and the physical meaning of each of the actions for each of the four orbit classes. The numbers in brackets after the orbit class are the orbit classification numbers used in Section~\ref{Sec::Accuracy}.}
\bgroup
\def\arraystretch{2.}
\begin{tabularx}{\columnwidth}{@{}XXXX@{}}
\toprule
Orbit class&$J_\lambda$&$J_\mu$&$J_\nu$\\
\midrule
Box (0)&\specialcell{$-\alpha\leq\lambda\leq\lambda_+$\\Length in $x$}&\specialcell{$-\beta\leq\mu\leq\mu_+$\\Length in $y$}&\specialcell{$-\gamma\leq\nu\leq\nu_+$\\Length in $z$}\\
Short-axis loop (1)&\specialcell{$\lambda_-\leq\lambda\leq\lambda_+$\\Radial extent}&\specialcell{$-\beta\leq\mu\leq-\alpha$\\Ang. mom.}&\specialcell{$-\gamma\leq\nu\leq\nu_+$\\Thickness in $z$}\\
Inner long-axis loop (2)&\specialcell{$-\alpha\leq\lambda\leq\lambda_+$\\Thickness in $x$}&\specialcell{$\mu_-\leq\mu\leq\mu_+$\\Radial extent}&\specialcell{$-\gamma\leq\nu\leq-\beta$\\Ang. mom.}\\
Outer long-axis loop (3)&\specialcell{$\lambda_-\leq\lambda\leq\lambda_+$\\Radial extent}&\specialcell{$\mu_-\leq\mu\leq-\alpha$\\Thickness in $x$}&\specialcell{$-\gamma\leq\nu\leq-\beta$\\Ang. mom.}\\
\bottomrule
\end{tabularx}
\egroup
\label{ActionMeaning}
\end{table}

The approach to finding the actions presented here requires an explicit form for $f$. In the next section we will show how we can circumnavigate the need for this explicit form, which allows us to use the same equations for a general potential.

\section{The Triaxial St\"ackel Fudge}\label{Sec::Triaxapproximation}
We now show how we can use the insights from St\"ackel potentials to estimate actions in a more general potential. For a general triaxial potential, $\Phi$, we can attempt to find the actions by assuming that the general potential is close to a St\"ackel potential. Given a general potential we define the quantities
\begin{equation}
\begin{split}
\chi_\lambda(\lambda,\mu,\nu) &\equiv (\lambda-\mu)(\nu-\lambda)\Phi(\lambda,\mu,\nu),\\
\chi_\mu(\lambda,\mu,\nu) &\equiv (\mu-\nu)(\lambda-\mu)\Phi(\lambda,\mu,\nu),\\
\chi_\nu(\lambda,\mu,\nu) &\equiv (\nu-\lambda)(\mu-\nu)\Phi(\lambda,\mu,\nu).
\end{split}
\end{equation}
where we have chosen a particular coordinate system, $(\alpha,\beta,\gamma)$ (see \S~\ref{Sec::CoordSysChoice}). If $\Phi$ were a St\"ackel potential, these quantities would be given by, for instance,
\begin{equation}
\chi_\lambda(\lambda,\mu,\nu) = f(\lambda)-\lambda\frac{f(\mu)-f(\nu)}{\mu-\nu}+\frac{\nu f(\mu)-\mu f(\nu)}{\mu-\nu}.
\end{equation}
Therefore, for a general potential, we can write
\begin{equation}
f(\tau) \approx \chi_\tau(\lambda,\mu,\nu)+C_\tau\tau+D_\tau,
\end{equation}
where $C_\tau$ and $D_\tau$ are constants provided we always evaluate $\chi_\tau$ with two of the ellipsoidal coordinates fixed. For instance, we always evaluate $\chi_\lambda$ at fixed $\mu$ and $\nu$.

When we substitute these expressions into equation~\eqref{Eq::EqnOfMotion} we find
\begin{equation}
2(\tau+\alpha)(\tau+\beta)(\tau+\gamma)p_\tau^2=\tau^2 E -\tau A_\tau+B_\tau +\chi_\tau(\lambda,\mu,\nu).
\label{Eq::EqnOfMotion_JK}
\end{equation}
For each $\tau$ coordinate there are two new integrals of motion given by $A_\tau=a-C_\tau$ and $B_\tau=b+D_\tau$.

Given an initial phase-space point, $(\vx_0,\vv_0)$, and a coordinate system, $(\alpha,\beta,\gamma)$, we can calculate the ellipsoidal coordinates $(\lambda_0,\mu_0,\nu_0,p_{\lambda 0},p_{\mu 0},p_{\nu 0})$. Inserting this initial phase-space point into equation~\eqref{Eq::EqnOfMotion_JK} gives us an expression for $B_\tau$ as
\begin{equation}
B_\tau = 2(\tau_0+\alpha)(\tau_0+\beta)(\tau_0+\gamma)p_{\tau 0}^2-\tau_0^2 E +\tau_0 A_\tau-  \chi_\tau(\lambda_0,\mu_0,\nu_0).
\label{Eq::K}
\end{equation}
It remains to find an expression for $A_\tau$ as a function of the initial phase-space point. To proceed we consider the derivative of the Hamiltonian with respect to $\tau$. In a St\"ackel potential we can stay on the orbit while changing $\tau$ and $p_\tau(\tau)$ with all the other
phase-space variables held constant. Therefore, in a St\"ackel potential $\upartial H/\upartial \tau = 0$. Here we consider $\upartial H/\upartial \lambda$ and will give the results for $\mu$ and $\nu$ afterwards. Using equation~\eqref{Eq::Hamiltonian} we write
\begin{equation}
\begin{split}
0&=\Big(\frac{\upartial H}{\upartial \lambda}\Big)_{\mu,\nu} \\
&= \half\frac{\upartial}{\upartial \lambda}\Big[\frac{p^2_\lambda}{P^2_\lambda}\Big]+\half\frac{p_\mu^2}{(\mu-\lambda)P_\mu^2}+\half\frac{p_\nu^2}{(\nu-\lambda)P_\nu^2}+\frac{\upartial \Phi}{\upartial \lambda}.
\label{Eq::Hderiv}
\end{split}
\end{equation}
To evaluate $\upartial[p^2_\lambda/P^2_\lambda]/\upartial \lambda $ we use
equation~\eqref{Eq::K} to write
\begin{equation}
\begin{split}
&2(\lambda+\alpha)(\lambda+\beta)(\lambda+\gamma)p_\lambda^2 = 2(\lambda_0+\alpha)(\lambda_0+\beta)(\lambda_0+\gamma)p_{\lambda 0}^2\\&+(\lambda^2-\lambda_0^2)E -(\lambda-\lambda_0)A_\lambda - \chi_\lambda(\lambda,\mu_0,\nu_0)+\chi_\lambda(\lambda_0,\mu_0,\nu_0),
\end{split}
\end{equation}
such that
\begin{equation}
\frac{p^2_\lambda}{P^2_\lambda} = \frac{Q+(\lambda^2-\lambda_0^2)E -(\lambda-\lambda_0)A_\lambda}{(\lambda-\mu)(\lambda-\nu)}-\Phi(\lambda,\mu_0,\nu_0),
\end{equation}
where
\begin{equation}
Q = 2(\lambda_0+\alpha)(\lambda_0+\beta)(\lambda_0+\gamma)p_{\lambda 0}^2+\chi_\lambda(\lambda_0,\mu_0,\nu_0).
\end{equation}
Upon substitution into equation~\eqref{Eq::Hderiv} we note that the derivatives of $\Phi$ cancel. Therefore, evaluating $\upartial H/\upartial \lambda$ at the initial phase-space point we find
\begin{equation}
\begin{split}
A_\lambda= & 2\lambda_0E-(2\lambda_0-\mu_0-\nu_0)\Big(\Phi(\lambda_0,\mu_0,\nu_0)+\half\frac{p_{\lambda 0}^2}{P_{\lambda 0}^2}\Big)\\&-\half\frac{p_{\mu 0}^2(\lambda_0-\nu_0)}{P_{\mu 0}^2}-\half\frac{p_{\nu 0}^2(\lambda_0-\mu_0)}{P_{\nu 0}^2}.
\end{split}
\end{equation}
This can be simplified further to
\begin{equation}
A_\lambda= (\mu_0+\nu_0)E+\half\frac{p_{\mu 0}^2(\lambda_0-\mu_0)}{P_{\mu 0}^2}+\half\frac{p_{\nu 0}^2(\lambda_0-\nu_0)}{P_{\nu 0}^2}.
\label{Eq::Jlam}
\end{equation}
Note that $A_\lambda$ is independent of $\lambda_0$ and $p_{\lambda 0}$ (except implicitly in the energy, $E$) as $P_{\tau 0}$ contains cancelling factors of $(\lambda_0-\tau_0)$.
Similarly
\begin{equation}
\begin{split}
A_\mu &= (\lambda_0+\nu_0)E+\half\frac{p_{\lambda 0}^2(\mu_0-\lambda_0)}{P_{\lambda 0}^2}+\half\frac{p_{\nu 0}^2(\mu_0-\nu_0)}{P_{\nu 0}^2},\\
A_\nu &= (\lambda_0+\mu_0)E+\half\frac{p_{\lambda 0}^2(\nu_0-\lambda_0)}{P_{\lambda 0}^2}+\half\frac{p_{\mu 0}^2(\nu_0-\mu_0)}{P_{\mu 0}^2}.
\end{split}
\label{Eq::Jmunu}
\end{equation}

For a \emph{true St\"ackel potential}, given an initial phase-space point we can
find 6 integrals of motion, $(A_\lambda,A_\mu,A_\nu,B_\lambda,B_\mu,B_\nu)$
from equations~\eqref{Eq::K},~\eqref{Eq::Jlam} and ~\eqref{Eq::Jmunu}. Note that a general St\"ackel potential only admits three integrals of motion so the 6 derived integrals of motion are not independent. This
procedure gives identical results to evaluating the integrals as in
\cite{deZeeuw1985}. Note that the expressions for these integrals do not
explicitly involve the function $f(\tau)$ -- they only involve the potential,
$\Phi$. With the integrals of motion calculated we are in a position to find
$p_\tau(\tau)$ and hence the actions from equation~\eqref{Eq::action}.

For a \emph{general potential} we may find six approximate integrals of motion using
the same equations, and hence estimate the actions. In this case, although the potential may admit only three true integrals of motion, the $6$ approximate integrals of motion are independent estimates of true integrals of motion. Again, as the expressions
do not require $f(\tau)$ they can be evaluated for a general potential. In Appendix~\ref{Appendix::Angles} we show how the angles and frequencies can be estimated using the same approach.

\subsection{Relation to axisymmetric case}\label{Sec::axisym}
The above procedure extends the work of \cite{Binney2012}. \cite{Binney2012} constructed the ``St\"ackel fudge'' algorithm for estimating actions in a general axisymmetric potential $\Phi(R,z)$, where $R$ and $z$ are the usual cylindrical polar coordinates. We now relate the procedure to that of \cite{Binney2012} to develop further understanding.

Oblate axisymmetric St\"ackel potentials are associated with prolate elliptic coordinates $(\lambda,\nu)$ given by the roots for $\tau$ of
\begin{equation}
\frac{R^2}{\tau+\alpha}+\frac{z^2}{\tau+\gamma} = 1,
\end{equation}
where $-\gamma\leq\nu\leq-\alpha\leq\lambda$. \cite{Binney2012} uses the coordinates $(u,v)$ which are related to $(\lambda,\nu)$ via
\begin{equation}
\begin{split}
\sinh^2 u &= \frac{\lambda+\alpha}{\gamma-\alpha},\\
\cos^2 v &= \frac{\nu+\gamma}{\gamma-\alpha},
\end{split}
\end{equation}
such that
\begin{equation}
\begin{split}
R &= \sqrt{\gamma-\alpha}\sinh u\sin v,\\
z &= \sqrt{\gamma-\alpha}\cosh u\cos v.
\end{split}
\end{equation}

An oblate axisymmetric St\"ackel potential can be written as
\begin{equation}
\Phi_S(\lambda,\nu) = -\frac{f(\lambda)-f(\nu)}{\lambda-\nu},
\end{equation}
and the equations for the momenta are given by \citep{deZeeuw1985}
\begin{equation}
2(\tau+\alpha)(\tau+\gamma)p_\tau^2 = E(\tau+\gamma)-\Big(\frac{\tau+\gamma}{\tau+\alpha}\Big)I_2-I_3+f(\tau).
\end{equation}
For axisymmetric potentials $I_2=\half L_z^2$, where $L_z$ is the $z$-component of the angular momentum. For a general oblate axisymmetric potential, $\Phi$, we define
\begin{equation}
\begin{split}
\chi_\lambda(\lambda,\nu) &\equiv -(\lambda-\nu)\Phi,\\
\chi_\nu(\lambda,\nu) &\equiv -(\nu-\lambda)\Phi.
\end{split}
\end{equation}
If $\Phi$ were a St\"ackel potential these quantities would be given by
\begin{equation}
\begin{split}
\chi_\lambda(\lambda,\nu) &= f(\lambda)-f(\nu),\\
\chi_\nu(\lambda,\nu) &= f(\nu)-f(\lambda).
\end{split}
\end{equation}
Therefore, for a general potential, we can write,
\begin{equation}
f(\tau) \approx \chi_\tau(\lambda,\nu)+D_\tau,
\end{equation}
where $D_\tau$ are constants provided we evaluate $\chi_\lambda$ at constant $\nu$ and vice versa. We can write the equations for the momenta as
\begin{equation}
2(\tau+\alpha)(\tau+\gamma)p_\tau^2 = E(\tau+\gamma)-\Big(\frac{\tau+\gamma}{\tau+\alpha}\Big)I_2-B_\tau+\chi_\tau(\lambda,\nu),
\end{equation}
where we have defined the integral of motion $B_\tau = I_3-D_\tau$. $B_\tau$ may be found given an initial phase-space point and we then integrate the equations for the momenta to find the actions. Note that in this case only two integrals of the motion, $B_\tau$, need to be found, as, in the axisymmetric case, we can find two exact integrals of motion, $E$ and $L_z$. This is the procedure followed in \cite{Binney2012} and, despite the differing conventions and presentation, this method gives identical results to that of \cite{Binney2012}.

\section{Tests}\label{Sec::Tests}
\begin{figure}
$$\includegraphics[width=\columnwidth,bb = 7 8 297 155]{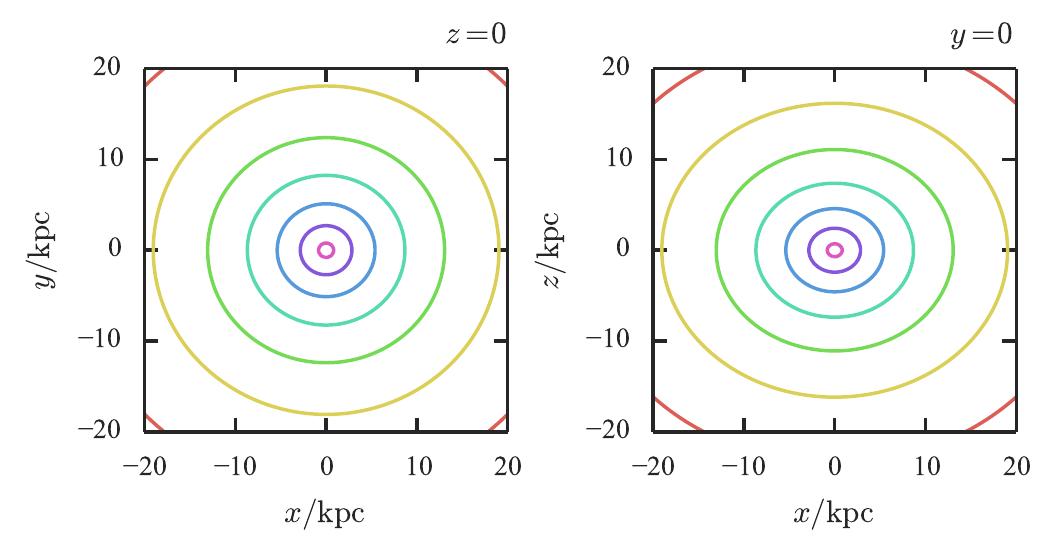}$$
\caption[Equipotential contours for the triaxial NFW potential]{Equipotential contours for the triaxial NFW potential in the two planes $z=0$ (left) and $y=0$ (right). The central contour shows $\Phi/GM_sm_0 = -0.0096$ and the contours increase linearly by $\Delta(\Phi/GM_sm_0) = 0.0008$ outwards.}
 \label{EquiPots}
 \end{figure}

For the purposes of testing the above algorithm, we use a triaxial NFW halo \citep{NFW,JingSuto2002}:
\begin{equation}
\begin{split}
\Phi(x,y,z)=\Phi(m) &= \frac{-G M_s}{m}\log\Big(1+\frac{m}{m_0}\Big)\\
&\text{where } m=\sqrt{x^2+\frac{y^2}{y_s^2}+\frac{z^2}{z_s^2}}.
\end{split}
\label{NFW_potential}
\end{equation}
We set $y_s=0.95$, $z_s=0.85$, $m_0=10\kpc$ and $GM_s=(1109\kms)^2\kpc$. In Fig.~\ref{EquiPots} we show the equipotential contours in the $z=0$ and $y=0$ planes. It is perhaps more conventional to include the triaxiality in the density \citep[e.g.][]{JingSuto2002}, but, for simplicity, we have chosen to include triaxiality in the potential. For our choice of parameters this leads to negative densities along the $z$-axis for $z\gtrsim130\kpc$. This is well outside the region we will probe in our experiments so we are not concerned that our model is unphysical at large $z$.

\subsection{Selection of coordinate system}\label{Sec::CoordSysChoice}
The accuracy of the above routine for a general potential will depend upon our choice of coordinate system, $(\alpha,\beta,\gamma)$. Note that the potential is fixed and this coordinate system acts only as a set of parameters in the algorithm to find the actions. We can freely set $\gamma=-1\kpc^2$ as the coordinate system only depends on $\Delta_1=\sqrt{\beta-\alpha}$ and $\Delta_2=\sqrt{\gamma-\beta}$. For each orbit we consider we are in a position to choose different $\Delta_i$. Here we consider how we can choose suitable $\Delta_i$ given an initial phase-space point.

In \cite{Sanders2012} the mixed derivative
$\upartial_\lambda\upartial_\nu[(\lambda-\nu)\Phi]$ was used to select an appropriate
coordinate system in an axisymmetric potential. For the triaxial case we
could construct a similar quantity:
$\upartial_\lambda\upartial_\mu\upartial_\nu[(\lambda-\mu)(\mu-\nu)(\nu-\lambda)\Phi]$.
However, this expression would involve third derivatives of the potential so
is undesirable. \cite{Binney2014} selected a coordinate system by fitting
ellipses to shell orbits at each energy, $E$. We follow a similar procedure:
we assume that the best choice of coordinate system is solely a function of
$E$.

In a St\"ackel potential the short-axis closed loops are ellipses confined to the plane $z=0$ with foci at $y=\pm\Delta_1=\pm\sqrt{\alpha-\beta}$, whilst the long-axis closed loops are confined to the plane $x=0$ with foci at $z=\pm\Delta_2=\pm\sqrt{\gamma-\beta}$. Additionally, for these closed loop orbits only one of the actions is non-zero ($J_\mu$ for the short-axis closed loop and $J_\nu$ for the long-axis closed loop).

For a general potential we use these facts to select appropriate values for
$\Delta_i$ using a two step procedure: given a value for $E$ we find the two closed loop orbits -- one
around the short axis and one around the long axis, and with these closed orbits found we alter the position of the foci to optimise the action estimates from our algorithm. Note that the structure of the closed orbits is independent of any choice of the foci positions such that the two steps of the procedure are distinct.

First, to find the closed orbits with energy $E$, we select a point along the intermediate axis, $y=y_I$, and launch an orbit with speed
$v=\sqrt{2(E-\Phi(0,y_I,0))}$ in either the $x$ (for the short-axis loop) or
$z$ direction (for the long-axis loop). The next time the orbit crosses the
$y$-axis we note the $y$-intercept, $y=y_F$ and calculate $|-y_F-y_I|$. We repeat this procedure with a new $y_I$ until we have minimised $|-y_F-y_I|$ using Brent's method. We only integrate half of
the orbit and assume that the other half can be obtained by symmetry to avoid
misidentifying fish-tail resonant orbits as closed loop orbits.

With the closed orbits with energy $E$ in our potential found, we turn to estimating the location of the foci. Using the long-axis closed loop orbit integration we find an estimate of $\Delta_2$ by minimising the standard deviation of the $J_\nu$ estimates from each time-step with respect to $\beta$ using Brent's method. The action estimates are found using the algorithm outlined in Section~\ref{Sec::Triaxapproximation}. This procedure is not sensitive to the choice of $\alpha$. Once we have found $\beta$ we perform a similar procedure for the short-axis loop: vary $\alpha$ until we have minimised the standard deviation of $J_\mu$.

\begin{figure}
$$\includegraphics[bb = 7 8 230 237]{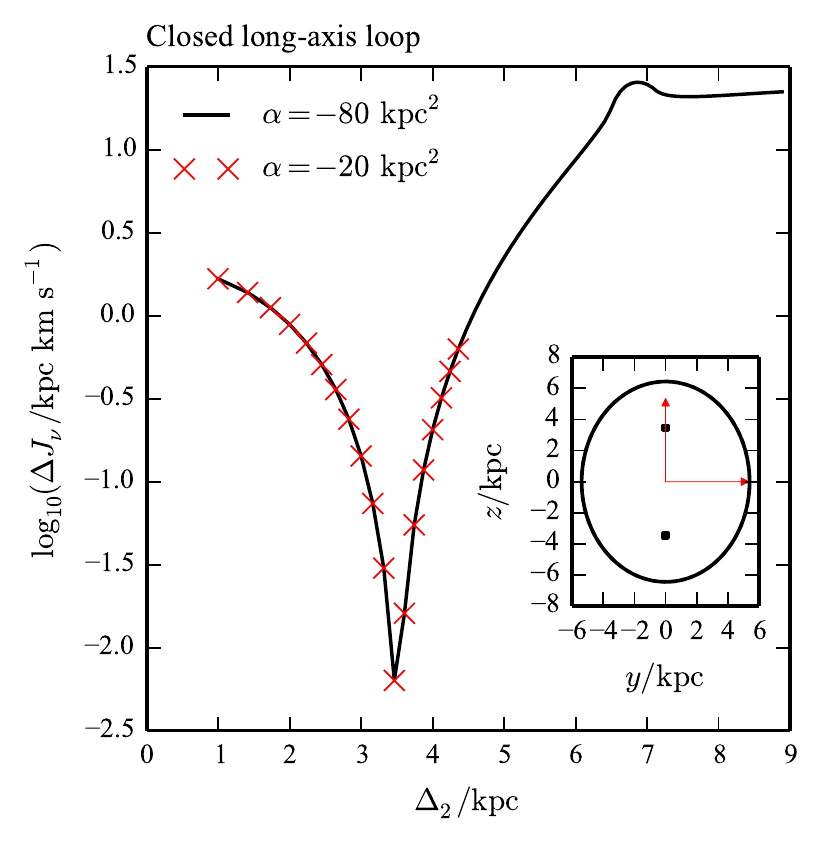}$$
\caption[Standard deviation in $J_\nu$ as a function of $\Delta_2$ for the closed long-axis loop orbit]{Standard deviation in $J_\nu$ as a function of $\Delta_2$ for the closed long-axis loop orbit shown in the inset. The solid line shows the results if we set $\alpha=-80\kpc^2$ whilst the red crosses show the results if we set $\alpha=-20\kpc^2$. The choice of $\Delta_2$ is insensitive to $\alpha$. In the inset, the two red arrows show the initial position vector for the orbit and that position vector rotated by $90\degrees$ anticlockwise. The black squares show the chosen location of the foci $z = \pm\Delta_2$.}
\label{Delta2}
\end{figure}

\begin{figure}
$$\includegraphics[bb = 7 8 225 237]{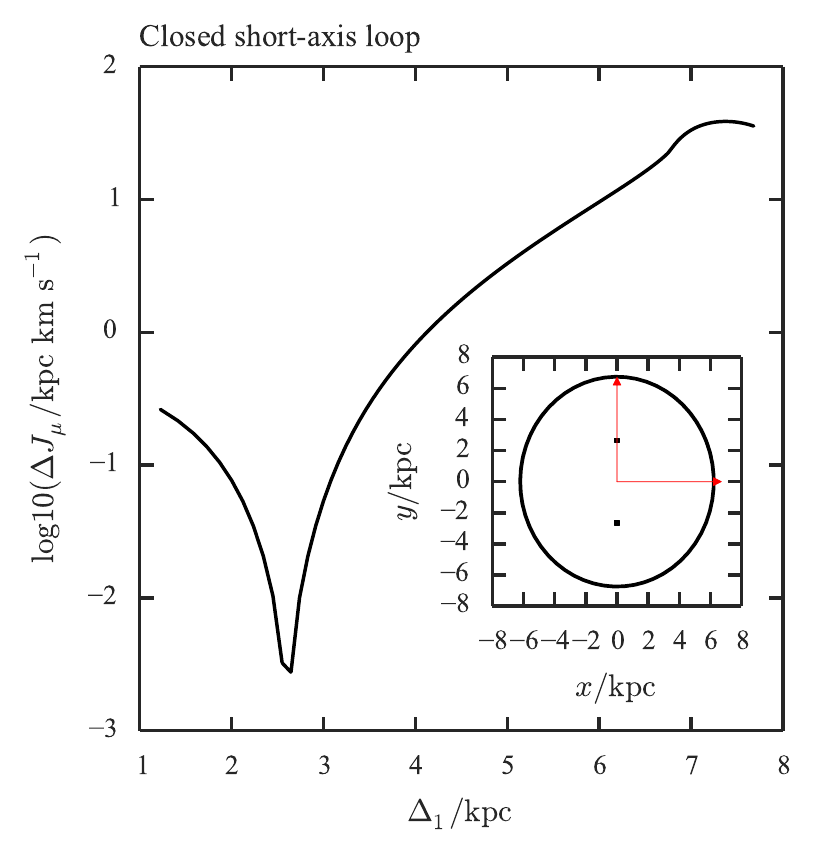}$$
\caption[Standard deviation in $J_\mu$ as a function of $\Delta_1$ for the closed short-axis loop orbit]{Standard deviation in $J_\mu$ as a function of $\Delta_1$ for the closed short-axis loop orbit shown in the inset. In the inset, the two red arrows show the initial position vector for the orbit and that position vector rotated by $90\degrees$ clockwise. The black squares show the chosen location of the foci $y=\pm\Delta_1$.}
\label{Delta1}
\end{figure}

We perform the above procedure for a range of energies from $E_{\rm min}=\Phi(0,y_{\rm min},0)$ to $E_{\rm max}=\Phi(0,y_{\rm max},0)$, tabulating the found values of $\alpha$ and $\beta$ for interpolation. For the NFW potential, we adopt $y_{\rm min}=0.05\kpc$ and $y_{\rm max}=60\kpc$. In Figs.~\ref{Delta2} and~\ref{Delta1} we plot the standard deviation of the actions of the closed loop orbits against $\Delta_2$ and $\Delta_1$ for the constant energy surface with $E=\Phi(0,m_0,0)=-(290\kms)^2$. In both cases there is a clear minimum in the standard deviation. In Fig.~\ref{Delta2} we show the standard deviation in $J_\nu$ as a function of $\Delta_2=\sqrt{\gamma-\beta}$ using two different values for $\alpha$. The results are indistinguishable. Provided we initially choose a sufficiently negative value of $\alpha$ that the optimal $\beta$ satisfies $\beta>\alpha$, we are free to first set $\Delta_2$ and then choose $\Delta_1$.

\subsubsection{Coordinate system procedure}
For clarity, we now summarise the above procedure:
\begin{enumerate}
\item Given a general potential, create a regularly-spaced grid in energy, $E$, between some minimum and maximum energy.
\item At each grid-point, $E_i$, find the short-axis and long-axis closed loops by integrating orbits launched at $(0,y_k,0)$ with velocity $\sqrt{2(E_i-\Phi(0,y_k,0))}$ in the direction of the long-axis or short-axis respectively. The closed loops will cross the $y$-axis for the first time at $(0,-y_k,0)$. We store the phase-space points $(\bs{x}_j,\bs{v}_j)$ at each time sample $t_j$.
\item Minimise the standard deviation of the $J_\nu(\bs{x}_j,\bs{v}_j)$ from the long-axis closed loop orbit integration with respect to $\beta$ to find $\Delta_2$.
\item Minimise the standard deviation of the $J_\mu(\bs{x}_j,\bs{v}_j)$ from the short-axis closed loop orbit integration with respect to $\alpha$ to find $\Delta_1$.
\end{enumerate}
We call the $\Delta_1$ and $\Delta_2$ found using this procedure the \emph{closed-loop}
estimates.

\subsubsection{Coordinate system results}

\begin{figure}
$$\includegraphics[bb = 7 8 220 173]{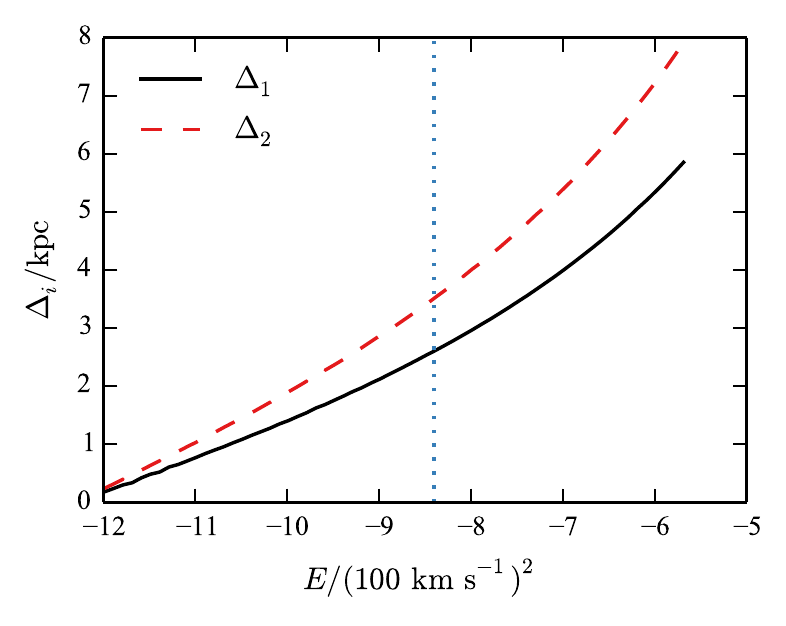}$$
\caption{Closed-loop choice of $\Delta_1$ (solid black) and $\Delta_2$ (dashed red) as a function of energy, $E$, for the NFW potential described in Section~\ref{Sec::Tests}. The range of energies covered corresponds to the energies of particles dropped from $0.5\kpc$ to $30\kpc$ along the intermediate axis. The vertical blue dotted line gives the energy of the surface explored in Section~\ref{Sec::Accuracy}.}
\label{Delta_withE}
\end{figure}

In Fig.~\ref{Delta_withE} we have plotted the closed-loop choice of $\Delta_1$ and $\Delta_2$ as a function of the energy. We see that for low energies (very centrally confined orbits) $\Delta_i$ tends to zero. Due to the cusp at the centre of the NFW potential, loop orbits exist right down to the centre of the potential. The foci must lie within these loop orbits so $\Delta_i$ must decrease as we go to lower energy. As we increase the energy $\Delta_i$ increases with $\Delta_1<\Delta_2$.

\begin{figure}
$$\includegraphics[width=\columnwidth,bb = 7 8 267 171]{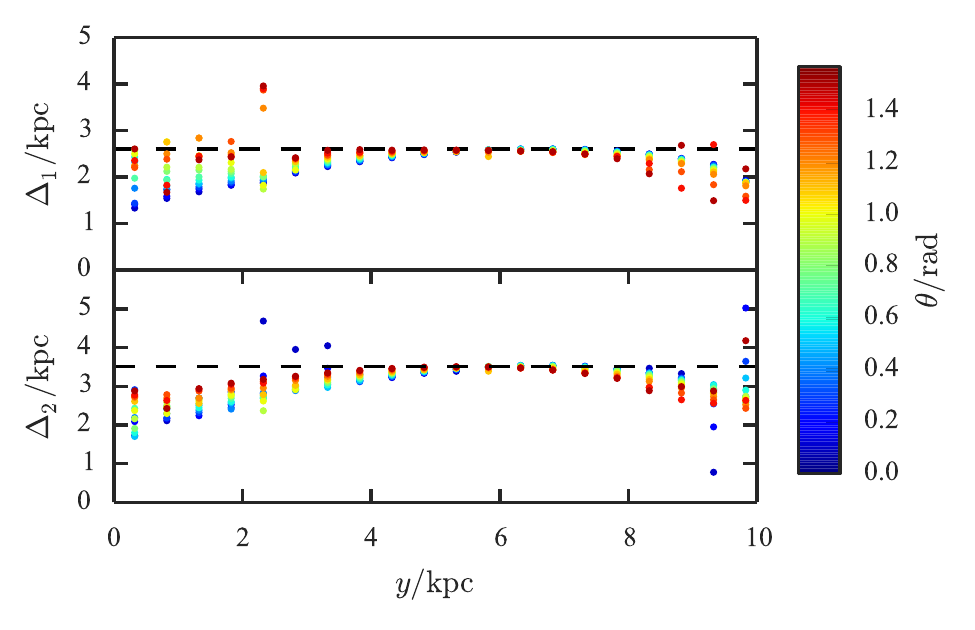}$$
\caption[Choice of $\Delta_1$ and $\Delta_2$ which minimise standard deviation of actions]{Choice of $\Delta_1$ and $\Delta_2$ which minimises the variation in the actions for a range of orbits confined to a constant energy surface. Each orbit was launched at $y$ on the intermediate axis with angle $\theta$ from the long axis. The dashed black line gives the values chosen by only  inspecting the closed loop orbits as specified in Section~\ref{Sec::CoordSysChoice}.}
\label{DeltaCheck}
\end{figure}

To check the closed-loop estimates we launch a series of orbits of constant energy $E =
\Phi(0,m_0,0)=-(290\kms)^2$ at linearly-spaced intervals along the $y$-axis
with velocity vectors in the $(x,z)$ plane oriented at differing
linearly-spaced angles, $\theta$, to the $x$ axis and integrate the orbits for approximately $10.3\Gyr$ storing phase-space points every $0.1\Gyr$. Note again that the orbit integration is in the fixed NFW potential and so the structure of an orbit is independent of any choice of $\alpha$ and $\beta$. The choice of $\alpha$ and $\beta$ only affects the recovery of the actions and we wish to find the optimal choice of $\alpha$ and $\beta$ for each orbit i.e. the choice that makes the actions as constant in time as possible. Therefore, we minimise the sum of the variances of the actions with respect to
$\alpha$ and $\beta$. The results of this procedure are shown in
Fig.~\ref{DeltaCheck}. We see that the majority of orbits yield optimal
$\Delta_i$ similar to the closed-loop estimates. At the extremes of $y$
$\Delta_i$ deviates from this choice. These are the box orbits and they seem
to favour lower $\Delta_i$. At fixed $y$ the choice of $\Delta_i$ is not so
sensitive to $\theta$.

We could improve our choice of $\Delta_1$ and $\Delta_2$ by making the choice
a function of an additional variable. For instance, we could make the choice
a function of the total angular momentum, which is not an integral of motion.
However, we will see that we cannot significantly improve the action recovery
with a better choice of $\Delta_i$.

\
\subsection{Accuracy}\label{Sec::Accuracy}

\begin{figure*}
$$\includegraphics[width=0.7\textwidth]{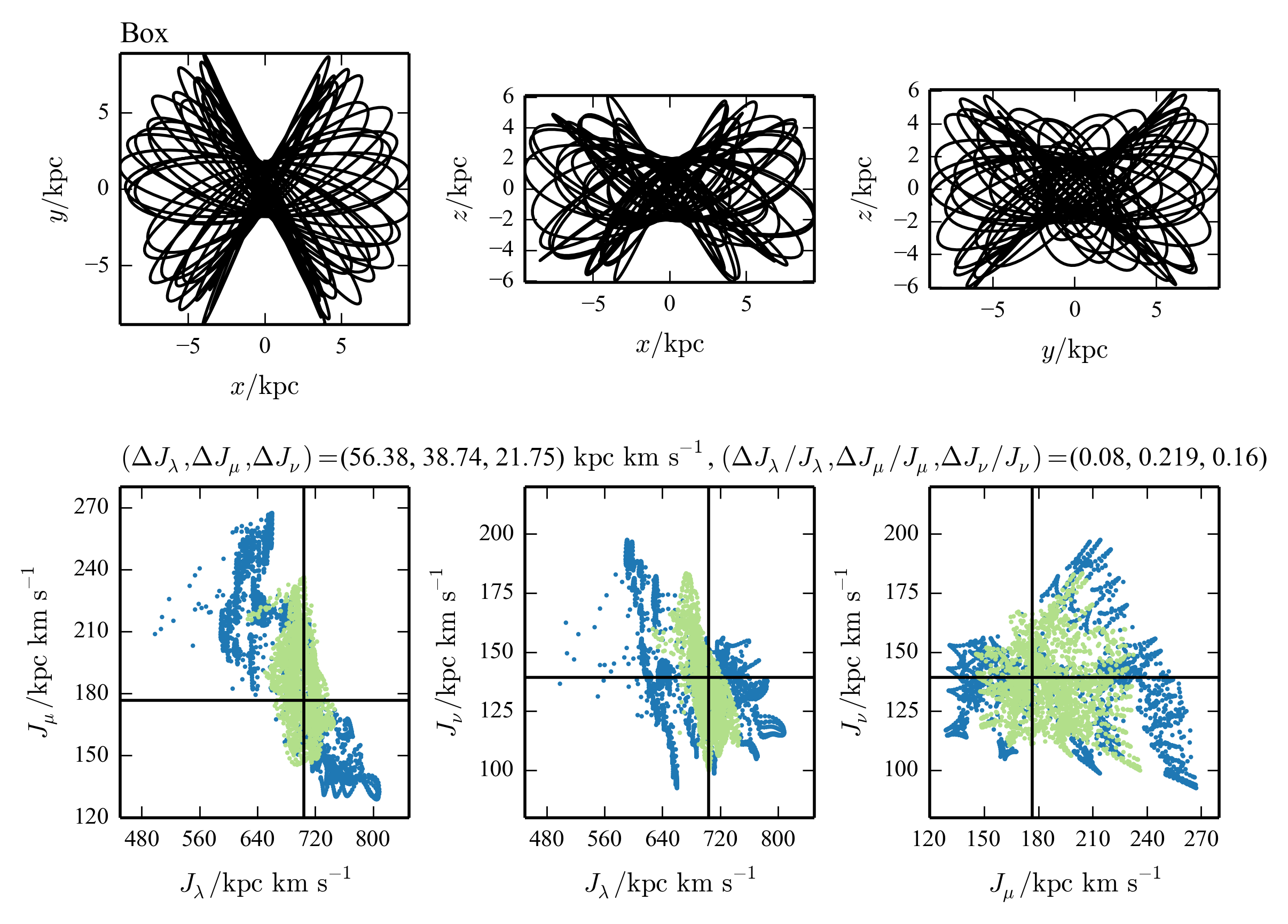}$$
\caption[Action estimates for an example box orbit using the triaxial St\"ackel fudge]{Action estimates for example box orbit using the triaxial St\"ackel fudge: the top three panels show three projections of the orbit, and the bottom three panels show the action estimates for points along the orbit. The dark blue points show the action estimates calculated using our closed-loop estimate of $\Delta_i$ based on the energy, the light green points show the choice of $\Delta_i$ that minimises the spread in the action estimates, and the black lines show the `true' actions found using the method presented by \protect\cite{SandersBinney2014}. Note that the origin is not included in the plots. Between the top and bottom plots, we give the absolute and relative error in the actions.}
\label{Box}
\end{figure*}

\begin{figure*}
$$\includegraphics[width=0.7\textwidth]{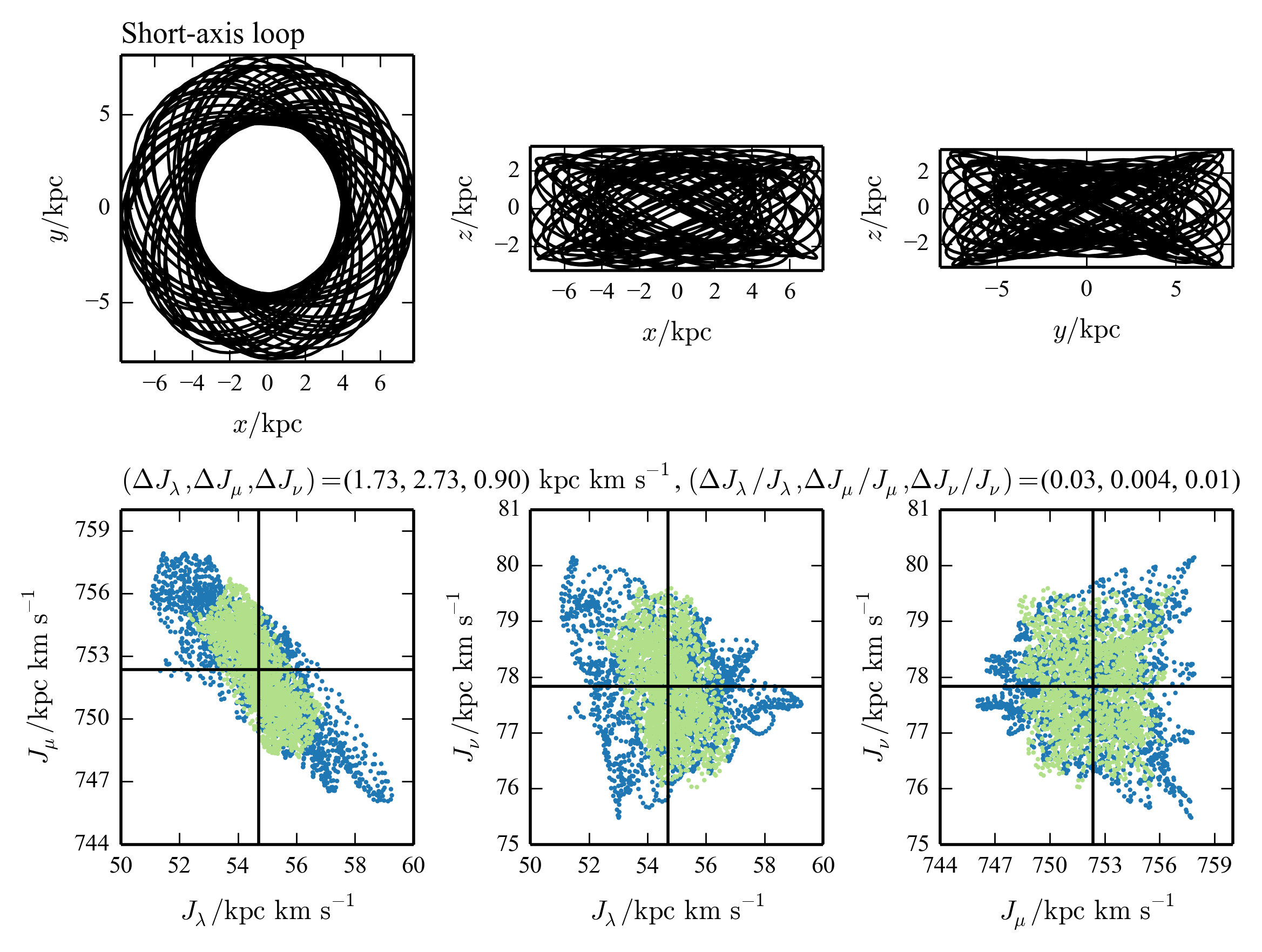}$$
\caption[Action estimates for an example short-axis loop orbit using the triaxial St\"ackel fudge]{Action estimates for example short-axis loop orbit using the triaxial St\"ackel fudge. See Fig.~\protect\ref{Box} for more information on each panel.
 }
\label{Short}
\end{figure*}

\begin{figure*}
$$\includegraphics[width=0.7\textwidth]{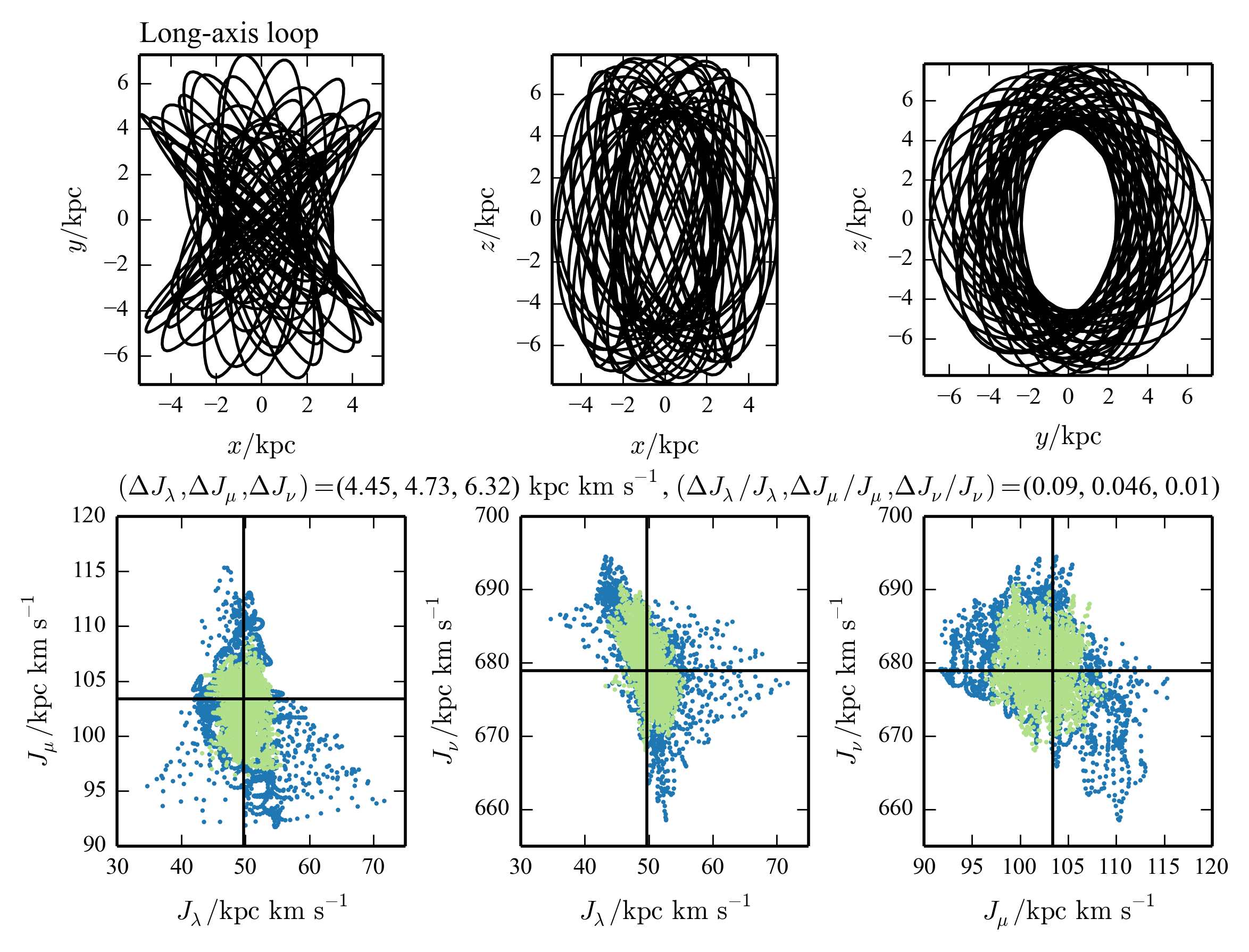}$$
\caption[Action estimates for an example long-axis loop orbit using the triaxial St\"ackel fudge]{Action estimates for example long-axis loop orbit using the triaxial St\"ackel fudge. See Fig.~\protect\ref{Box} for more information on each panel.
}
\label{Long}
\end{figure*}

We now briefly inspect the accuracy of the action recovery using the triaxial
St\"ackel fudge. We take three orbits from the surface of constant energy
explored in the previous section. The three orbits are a box orbit with
$y=1.8234\kpc$, $\theta=0.6\rad$ (shown in Fig.~\ref{Box}), a short-axis loop
orbit with $y=4.8234\kpc$, $\theta=0.4\rad$ (shown in Fig.~\ref{Short}), and
a long-axis loop orbit with $y=3.8234\kpc$, $\theta=1.2\rad$ (shown in
Fig.~\ref{Long}). The top row of each figure shows three projections of the
orbit, while the three lower panels show the action estimates calculated at
each point along the orbit using the closed-loop choice of $\Delta_i$ in
blue, and in green those obtained with the choice of $\Delta_i$ that
minimises the spread in the action estimates.  Clearly no procedure for
determining $\Delta_i$ will give superior performance to that obtained with
the latter, which is expensive to compute because it requires orbit
integration.  The intersection of the black lines in the bottom panels of
Figs.~\ref{Box} to \ref{Long} show the `true' actions calculated with the
method of \cite{SandersBinney2014}. The distributions of coloured points from
the St\"ackel Fudge scatter around the true actions, as one would hope. The
extent of the green distributions, obtained with the computationally costly
values of $\Delta_i$, are at best a factor two smaller that the distributions
of blue points, obtained with the cheap value of $\Delta_i$. From this
experiment we conclude that there is not a great deal to be gained by
devising a better way to evaluate the $\Delta_i$.

In Appendix~\ref{Appendix::Angles} we show how well
the angle coordinates are recovered for these orbits.

The actions of the box orbit are
$(J_\lambda,J_\mu,J_\nu)=(686,192,137)\kpc\kms$ and our method yields errors
of $(\Delta J_\lambda,\Delta J_\mu,\Delta J_\nu)=(56,39,22)\kpc\kms$ so
approximately $10-20\percent$. If we adjust $\Delta_i$ to minimise the spread
in the action estimates along the orbit, we find errors of $(\Delta
J_\lambda,\Delta J_\mu,\Delta J_\nu)=(17,19,16)\kpc\kms$ so approximately
$\lesssim10\percent$. We can achieve a factor of two improvement for
$J_\lambda$ and $J_\mu$.

The actions of the short-axis loop orbit are
$(J_\lambda,J_\mu,J_\nu)=(55,752,78)\kpc\kms$ and our method yields errors of
$(\Delta J_\lambda,\Delta J_\mu,\Delta J_\nu)=(2,3,1)\kpc\kms$ so $\lesssim
4\percent$. If we adjust $\Delta_i$ to minimise the spread in the action
estimates along the orbit, we find errors of $(\Delta J_\lambda,\Delta
J_\mu,\Delta J_\nu)=(0.8,2.0,0.9)\kpc\kms$.

The actions of the long-axis loop orbit are
$(J_\lambda,J_\mu,J_\nu)=(50,102,680)\kpc\kms$ and our method yields errors
of $(\Delta J_\lambda,\Delta J_\mu,\Delta J_\nu)=(4,5,6)\kpc\kms$ so
$\lesssim 8\percent$. If we adjust $\Delta_i$ to minimise the spread in the
action estimates along the orbit, we yield errors of $(\Delta
J_\lambda,\Delta J_\mu,\Delta J_\nu)=(2.0,2.5,4.2)\kpc\kms$.

\begin{figure}
$$\includegraphics[bb = 7 8 230 358,width=0.45\textwidth]{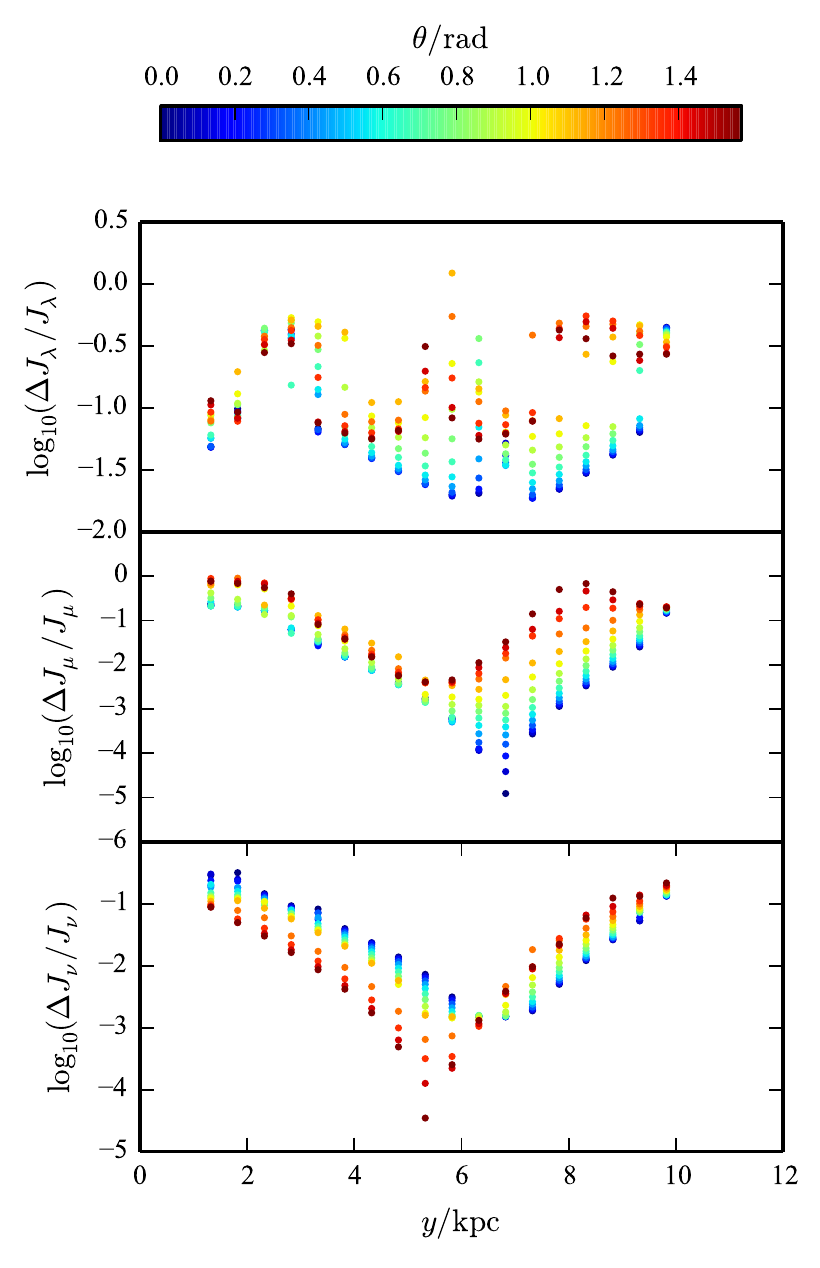}$$
\caption[Error in the actions for a selection of orbits]{Logarithm of the fractional error in the actions for a selection of orbits in the constant energy surface $E = \Phi(0,m_0,0)=-(290\kms)^2$ for the triaxial NFW potential. The $x$-axis shows the position along the intermediate axis at which the orbits were launched ($y$), and the colour-coding in the right panel shows the angle, $\theta$, in the $x-z$ plane at which the orbits were launched.}
\label{OrbitAcc10}
\end{figure}

For all the orbits shown in Fig.~\ref{DeltaCheck} (sampled from the constant
energy surface $E=\Phi(0,m_0,0)=-(290\kms)^2$), we have plotted the logarithm
of the fractional error in the actions in Fig.~\ref{OrbitAcc10} (i.e. logarithm of the standard deviation of the action estimates around the orbit over the mean action estimate). We find the
most accurate action recovery occurs for the orbits with the initial
condition $y\approx m_0/2$, where we have mostly loop orbits. For these loop
orbits, $J_\mu$ and $J_\nu$ are accurate to $\lesssim1\percent$ but the
`radial' action $J_\lambda$ is small for these orbits so the relative
error can be large. For the box orbits at the extremes of $y$, the
relative error increases to $\sim 10\percent$ but can be as large as order
one in $J_\mu$ for low $y$.

\begin{figure*}
$$\includegraphics[bb = 1 4 773 269,width=\textwidth]{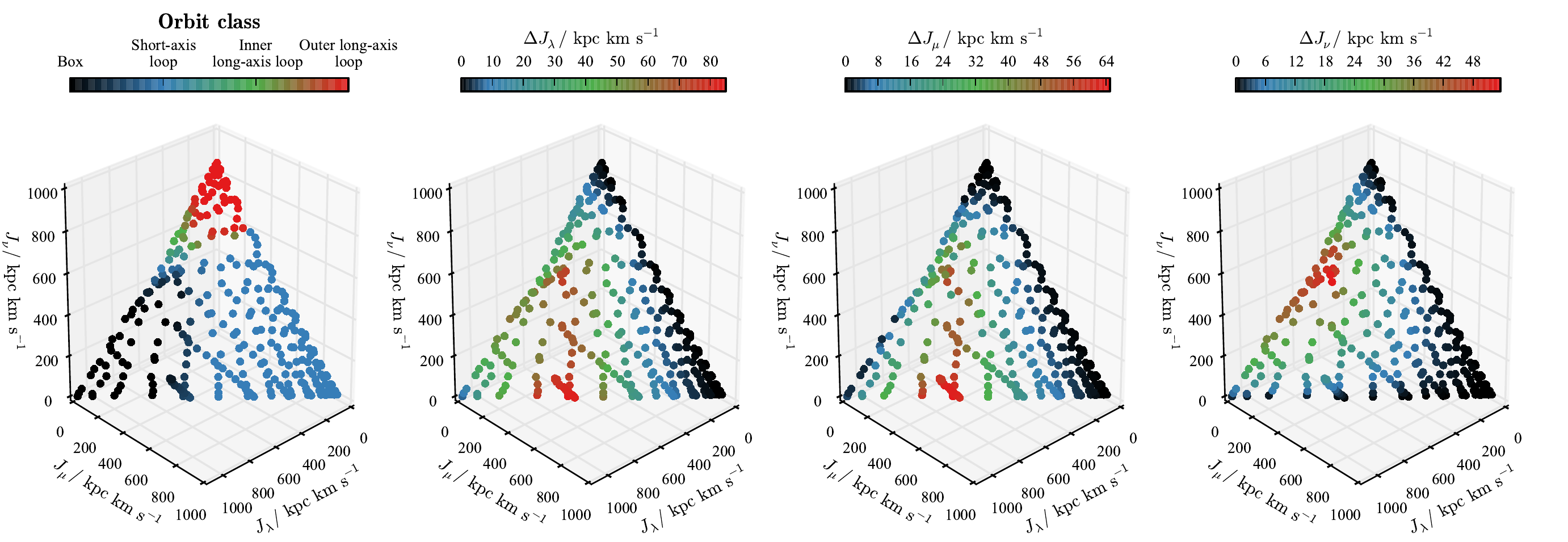}$$
\caption{Absolute errors in the actions as functions of action in the constant energy surface $E = \Phi(0,m_0,0)=-(290\kms)^2$ for the NFW potential. The leftmost panel shows the constant energy surface coloured by orbit class: boxes in black, short-axis loops in blue, inner long-axis loops in green, and outer long-axis loops in red. Note the classification is a continuum as it is calculated from an average of classifications along an orbit. The second, third and fourth panels show the absolute error in the three actions, $J_\lambda$, $J_\mu$ and $J_\nu$ respectively.}
\label{3Dactionerrors}
\end{figure*}

In Fig.~\ref{3Dactionerrors} we show the absolute errors in the actions as a function of action for the constant energy surface along with the orbit classification. These are again calculated as the standard deviation of action estimates around the orbit. Each phase-space point along the orbit is allocated a classification number based on the limits of $\tau$ found in the St\"ackel approximation (see Table~\ref{ActionMeaning}): $\lambda_-=-\alpha, \mu_-=-\beta$ and $\nu_-=-\gamma$ correspond to a box orbit (classification number 0), $\mu_-=\beta$, $\mu_+=-\alpha$ to a short-axis loop orbit (1), $\lambda_-=-\alpha$, $\nu_+=-\beta$ to an inner long-axis loop (2), and $\mu_+=-\alpha$, $\nu_+=-\beta$ to an outer long-axis loop (3). The orbit classification number is calculated as an average of these classifications along the orbit. With this scheme, orbits near the boundaries of the orbit classes that are chaotic or resonant are allocated non-integer orbit classification numbers. We see that the largest action errors occur at the interfaces between the orbit classes. In particular, $\Delta J_\lambda$ and $\Delta J_\mu$ are largest along the box-short-axis-loop interface, whilst $\Delta J_\nu$ is largest at the box-long-axis-loop interface. It is at these boundaries that the orbits pass close to the foci so clearly our choice of foci affects the action recovery for these orbits.

In general, we find that the action recovery for loop orbits is good, as these orbits probe a small radial range of the potential. For box orbits the recovery deteriorates as these orbits probe a larger central region of the potential. Additionally, we have seen that by altering $\Delta_i$ we can achieve up to a factor of two improvement in the accuracy of the actions for both the loop and box orbits.


\subsection{Surfaces of section}

\begin{figure*}
\begin{tabular}{lll}
$$\includegraphics[bb = 7 8 201 135, width=0.3\textwidth]{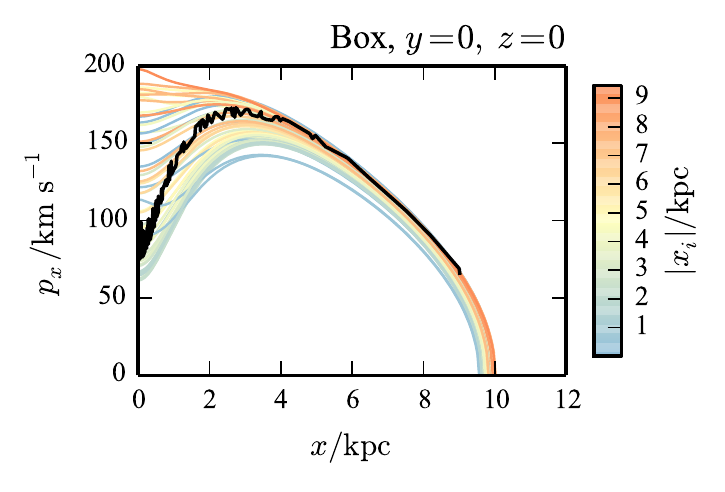}$$&
$$\includegraphics[bb = 7 8 203 135, width=0.3\textwidth]{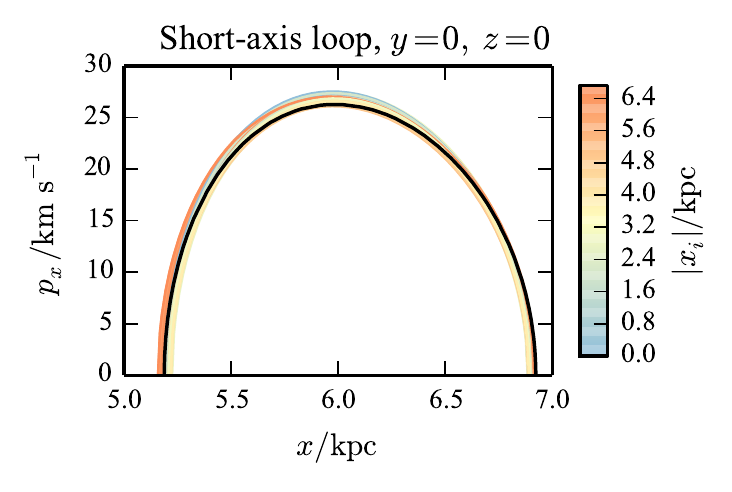}$$&
$$\includegraphics[bb = 7 8 204 135, width=0.3\textwidth]{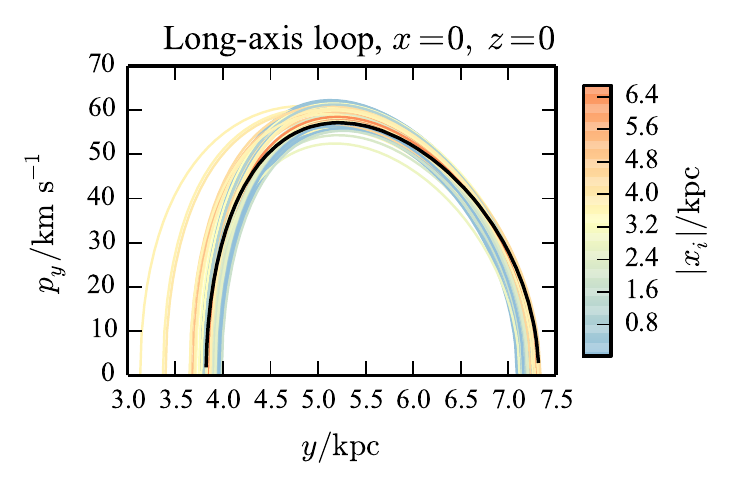}$$
\end{tabular}
\caption{Surfaces of section for the three test orbits in the triaxial NFW potential. In the left panel we show the box orbit, the central panel shows the short-axis loop orbit and the right panel shows the long-axis loop orbit. In each panel the solid black line gives the true curve of consequents found from orbit integration. The narrower coloured lines give the consequents from the St\"ackel approximation coloured by $|x_i|$ of the initial phase-space point where $x_i=x$ for the short-axis loop and box orbit, and $x_i=y$ for the long-axis loop orbit. The text above each plot gives the plane that defines the surface of section.}
\label{SoSfig}
\end{figure*}

For understanding the behaviour of dynamical systems, \cite{Poincare} introduced the concept of a surface of section. These diagrams simplify the motion of a high-dimensional dynamical system. A regular orbit in an integrable triaxial potential permits three constants of the motion, thus confining the motion to a 3-torus. If we choose to only plot the series of points where the orbit passes through a 4-surface in phase-space, e.g. defined by $y=0$ and $z=0$, the phase-space points will be confined to a line, or a consequent, which may be visualized clearly.

We can test the St\"ackel approach outlined here by seeing how well it reproduces the surfaces of section. To produce the true surface of section we must integrate the orbit in the true potential and find the phase-space points where the orbit crosses our chosen 4-surface. Here we use 4-surfaces defined by one of the spatial axes. The orbit will never pass through a given spatial axis in a finite time so we can only produce points arbitrarily close to the given axis. If we require the points where the orbit crosses the $x$-axis we integrate until the $y$ and $z$ steps have bracketed $y=0$ and $z=0$. We then bisect the integration step $N_{\rm max}$ times choosing the interval that brackets $y=0$ and if the interval still brackets $z=0$ after the $N_{\rm max}$ bisections we store the final point. We integrate over our chosen step-size with a Dortmund-Prince 8th order adaptive integration scheme with a absolute accuracy of $\epsilon=1\times10^{-10}$. We choose a step-size of $0.005\kpc$ and set $N_{\rm max}=10$. This scheme produces points that are $\lesssim 0.001\kpc$ away from the $x$ axis for the box orbit considered below.

To produce the corresponding surface of section from the St\"ackel method we determine $\tau$ along our chosen spatial axis using equation~\eqref{Eq::CartFromTau} between the determined limits in $\tau$, and use equation~\eqref{Eq::EqnOfMotion_JK} to find the corresponding $p_\tau$. From $p_\tau$, we can use expressions such as equation~\eqref{Eq::Example_ptau} to calculate $p_x$, $p_y$ and $p_z$: if we wish to draw the consequent defined by $y=0$, $z=0$ we have that $\mu=-\beta$ and $\nu=-\gamma$ such that $x=\sqrt{\lambda+\alpha}$, and $p_x = \sqrt{4(\lambda+\alpha)}p_\lambda$. If we wish to draw the consequent defined by $x=0$, $z=0$ we have that for $|y|>\Delta_1$, $\mu=-\alpha$ and $\nu=-\gamma$ such that $y=\sqrt{\lambda+\beta}$, and $p_y = \sqrt{4(\lambda+\beta)}p_\lambda$, whilst for $|y|<\Delta_1$, $\lambda=-\alpha$ and $\nu=-\gamma$ such that $y=\sqrt{\mu+\beta}$ and $p_y=\sqrt{4(\mu+\beta)}p_\mu$\footnote{If we wish to draw the consequent defined by $x=0$, $y=0$ we have for $|z|<\Delta_2$, $\lambda = -\alpha$ and $\mu=-\beta$ so $z=\sqrt{\nu+\gamma}$ and $p_z = \sqrt{4(\nu+\gamma)}p_\nu$, whilst for $|z|>\Delta_2$, $\mu=-\alpha$ and $\nu=-\beta$ so $z=\sqrt{\lambda+\gamma}$ and $p_z = \sqrt{4(\lambda+\gamma)}p_\lambda$.}.

Fig.~\ref{SoSfig} shows that the St\"ackel approximation consequents for the short-axis loop
lie close to the true consequent. Those of the long-axis loop are slightly worse.
The box orbit seems problematic. For the phase-space points that lie close to
the centre of the potential the consequents turn over in the centre as
required. However they underestimate $p_x$ at a given $x$. The phase-space
points which lie further out fail to turn over at low $x$. The St\"ackel tori
for these orbits are near radial such that $p_x$ is maximum for $x=0$.
However, we see from Fig.~\ref{Box} that the orbit crosses through $x=0,y=0$
at an angle such that $p_x$ is smaller than its maximum value. This behaviour
is only captured for the initial phase-space points at low $x$.

\section{A triaxial model with specified \df{}}\label{Sec::DF}
The main purpose of the algorithm presented here is to calculate efficiently the moments of triaxial distribution functions. We have seen that the errors in the actions reported by the scheme can be large. However, when calculating moments of a distribution function, many action evaluations are required and there is scope for errors to substantially cancel, leaving the final value of the moment quite accurate. In this section, we demonstrate this phenomenon by for the first time constructing triaxial models from an analytic \df{} $f(\vJ)$.

We adopt a simple distribution function \citep{Posti}
\begin{equation}
f(\vx,\vv) = f(\vJ(\vx,\vv)) = (J_0+|J_\lambda|+\zeta|J_\mu|+\eta|J_\nu|)^p,
\end{equation}
where $J_0=10\kms\kpc$ is a scale action, $\zeta$ controls whether the model is tangentially/radially biased and $\eta$ controls the flattening in $z$ of the model. We do not construct a self-consistent model but instead consider $f$ to be a tracer population in the externally applied triaxial NFW potential of equation~\eqref{NFW_potential}.

\begin{figure*}
$$\includegraphics[bb = 7 8 566 198, width=\textwidth]{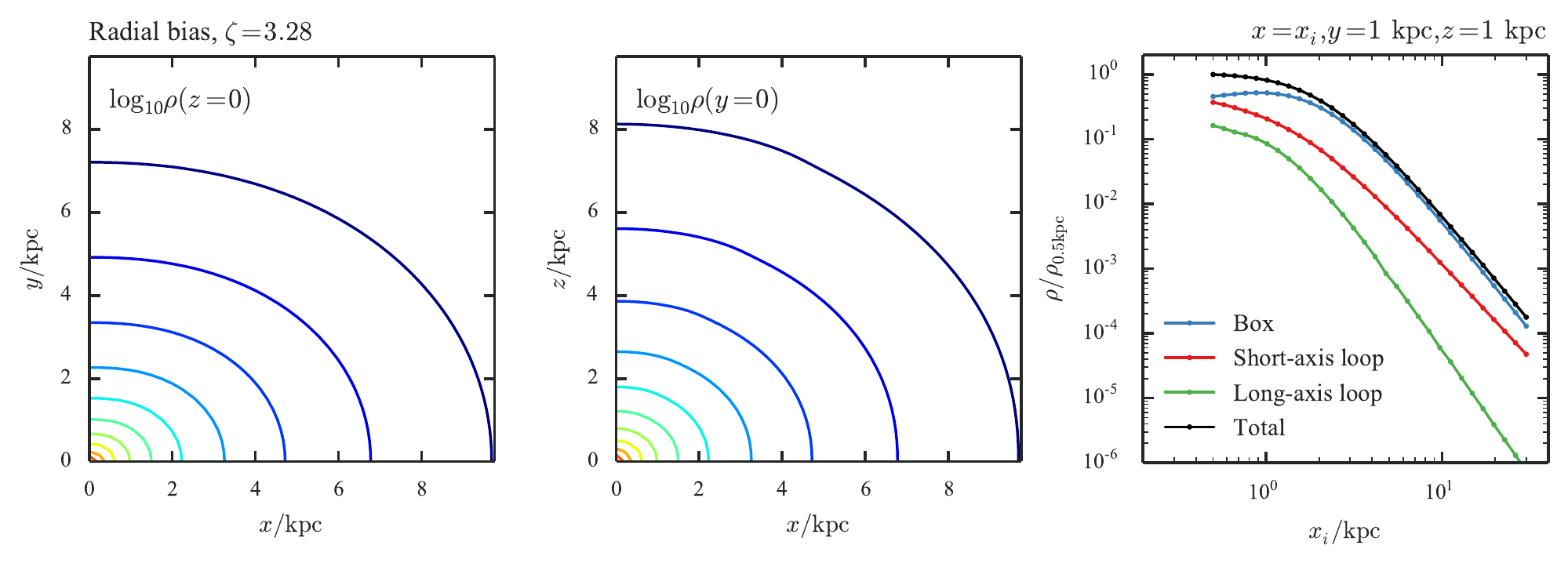}$$
\caption{Density for the radially-biased model ($\zeta=3.28$). The left panel shows equally-spaced contours of the logarithm of the density in the $(x,y)$ plane and similarly for the central panel in the $(x,z)$ plane. The outermost contour corresponds to $\log_{10}(\rho/\kpc^{-3})=0.5$ and the contours increase by $0.5$ inwards. The right panel shows the total density in black along the line $y=1\kpc, z=1\kpc$ as well as the contributions from the box orbits in blue, the short-axis loop orbits in red and the long-axis loop orbits in green.}
\label{Radial}
\end{figure*}

\begin{figure*}
$$\includegraphics[bb = 7 8 566 198, width=\textwidth]{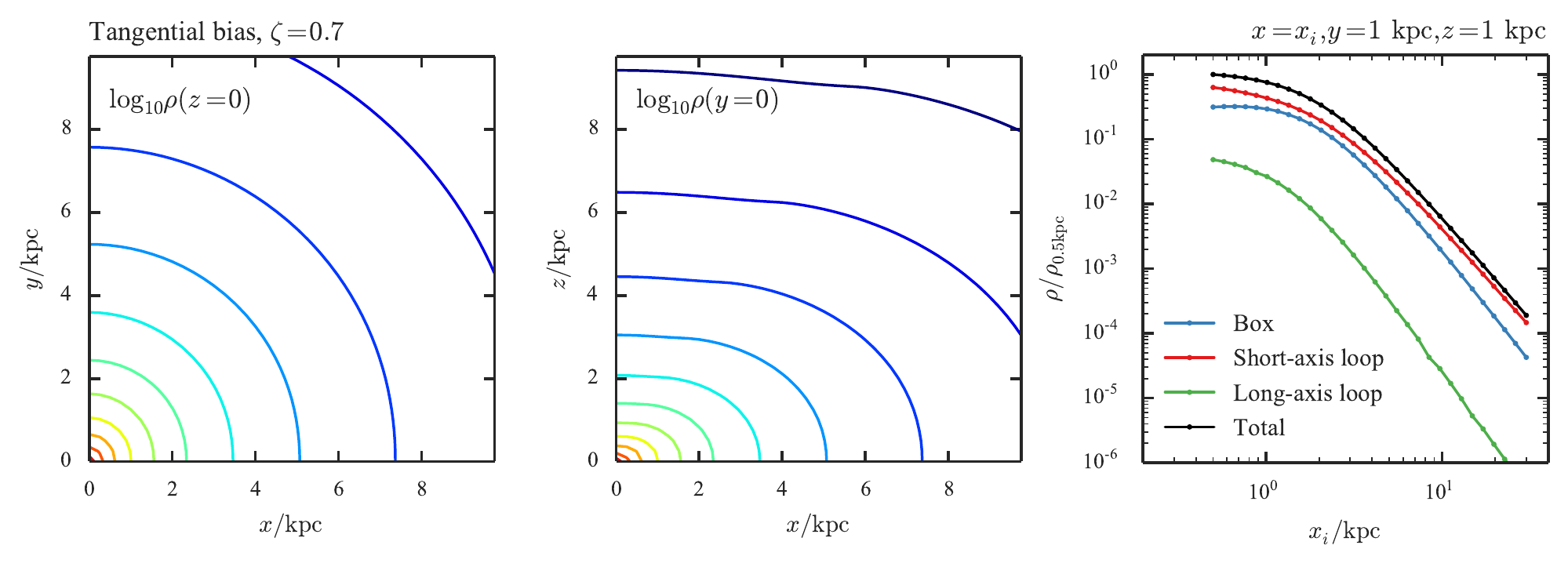}$$
\caption{Density for the tangentially-biased model ($\zeta=0.7$). The left panel shows equally-spaced contours of the logarithm of the density in the $(x,y)$ plane and similarly for the central panel in the $(x,z)$ plane. The outermost contour corresponds to $\log_{10}(\rho/\kpc^{-3})=1.0$ in the left panel and $\log_{10}(\rho/\kpc^{-3})=0.5$ in the central panel, and the contours increase by $0.5$ inwards. The right panel shows the total density in black along the line $y=1\kpc, z=1\kpc$ as well as the contributions from the box orbits in blue, the short-axis loop orbits in red and the long-axis loop orbits in green.}
\label{Tangential}
\end{figure*}

We set $p=-3$, which, for our choice of potential, causes the density to go as $r^0$ in the centre and fall off as $r^{-3}$ for large $r$. Note that the mass of this model diverges logarithmically. We set $\eta=1.88$ and explore two values of $\zeta=0.7$ (tangential bias) and $\zeta=3.28$ (radial bias). Note that for the orbit classes to fill action space seamlessly, we must scale the radial action of the loop orbits by a factor of two \citep{BinneySpergel}. We proceed by calculating the moments of this distribution function in the test triaxial NFW potential at given spatial points, $\vx$. These non-zero moments are
\begin{equation}
\begin{split}
\rho(\vx) &= \int\,\mathrm{d}^3\vv\,f(\vx,\vv),\\
\sigma_{ij}^2(\vx) &= \frac{1}{\rho}\int\,\mathrm{d}^3\vv\,v_i v_jf(\vx,\vv).
\end{split}
\end{equation}
Note that, as the potential is time-independent, the Hamiltonian is time-reversible and we need only integrate over half the velocity space and multiply the result by two. We integrate up to the maximum velocity, $v_{\rm max}$ at $\vx$, given by $v_{\rm max} = \sqrt{2\Phi(\vx)}$. We will later calculate these moments extensively to demonstrate that the action-based distribution functions obey the Jeans' equations. In Figs.~\ref{Radial} and~\ref{Tangential} we plot the density of the radially-biased ($\zeta=3.28$) and tangential-biased ($\zeta=0.7$) models. We display contours of constant density in two planes along with the density along a line parallel to the $x$-axis decomposed into its contributions from each orbit class. The density is calculated using the adaptive Monte-Carlo Divonne routine in the {\sc cuba} package of \cite{Hahn}. The class of each orbit is determined by the limits of the motion in $\tau$: $\lambda_-=-\alpha$, $\mu_-=-\beta$ and $\nu_-=-\gamma$ correspond to a box orbit, $\mu_-=\beta$, $\mu_+=-\alpha$ to a short-axis loop orbit, and $\nu_-=-\gamma$, $\nu_+=-\beta$ to a long-axis loop. As we are calculating the density close to the $x$-axis, the long-axis loop orbits, which loop the $x$-axis, do not contribute significantly to the density integral. We see that for the radially-biased model the box orbits are the dominant contributors whilst for the tangentially-biased model the short-axis loop orbits are the major contributors.

We will now perform some checks to see whether our distribution functions are accurate.

\subsection{Normalization}
One quick check of our action estimation scheme is how accurately it recovers the normalization. To keep the normalization finite we set $p=-3.5$ for this section. We are able to calculate the normalization of our \df{} in two distinct ways. Firstly, we calculate the normalization analytically from the \df{} as
\[
\begin{split}
M_{\rm true} &= (2\pi)^3\int\d^3\vJ\,f(\vJ) \\&= (2\pi)^3\int_0^\infty\d J_\lambda\,\int_0^\infty\d J_\mu\,\int_0^\infty\d J_\nu\,f(\vJ)\\
&=-\frac{(2\pi)^3J_0^{p+3}}{\eta\zeta(p+1)(p+2)(p+3)}.
\end{split}
\]
Note that for each $\vJ$ in the appropriate range there are two loop orbits
-- one circulating clockwise and one anti-clockwise. Therefore, we must
multiply the normalization by two for these orbits. However, we have defined
the `radial' action to be four times the integral from $\tau_-$ to $\tau_+$
for these orbits so these factors cancel \citep{BinneySpergel,deZeeuw1985}.
Additionally, we can calculate the mass as
\[
M_{\rm est} = 8\int_{(x,y,z)>0}\d^3\vx\,\int\d^3\vv\,f(\vJ(\vx,\vv)).
\]
For each spatial coordinate, we make the transformation $u_i=1/(1+x_i)$ to make the integrand flatter. The limits of the integral are now $u_i=[0,1]$. To reduce numerical noise, we split the integral such that we calculate the contribution near the axes separately. We perform the integral using the Monte Carlo Divonne routine. For the tangentially-biased model ($\zeta=0.7$), we find $M_{\rm est}\approx1.006M_{\rm true}$ and for the radially-biased model ($\zeta=3.28$) $M_{\rm est}\approx 1.007 M_{\rm true}$ so despite the often large errors in the actions the normalization of the model is well recovered.

\subsection{The Jeans equation}
Our distribution function must satisfy the collisionless Boltzmann equation
\begin{equation}
\frac{\d f}{\d t} = 0.
\end{equation}
In turn this means the distribution function must satisfy the Jeans equations (see equation (4.209) of \cite{BinneyTremaine})
\begin{equation}
\frac{\upartial(\rho \sigma_{ij}^2)}{\upartial x_i} = -\rho\frac{\upartial\Phi}{\upartial x_j}.
\end{equation}

\begin{figure*}
\begin{tabular}{lll}
$$\includegraphics[bb = 7 8 171 190, width=0.31\textwidth]{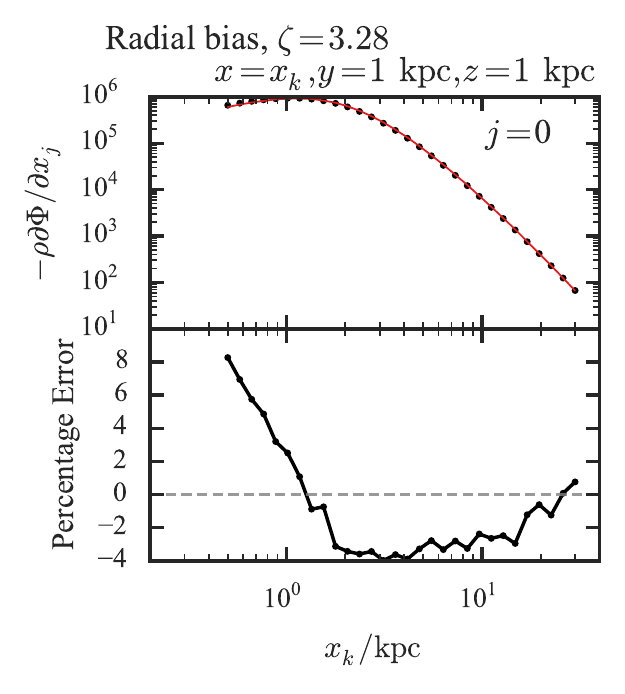}$$&
$$\includegraphics[bb = 7 8 171 181, width=0.31\textwidth]{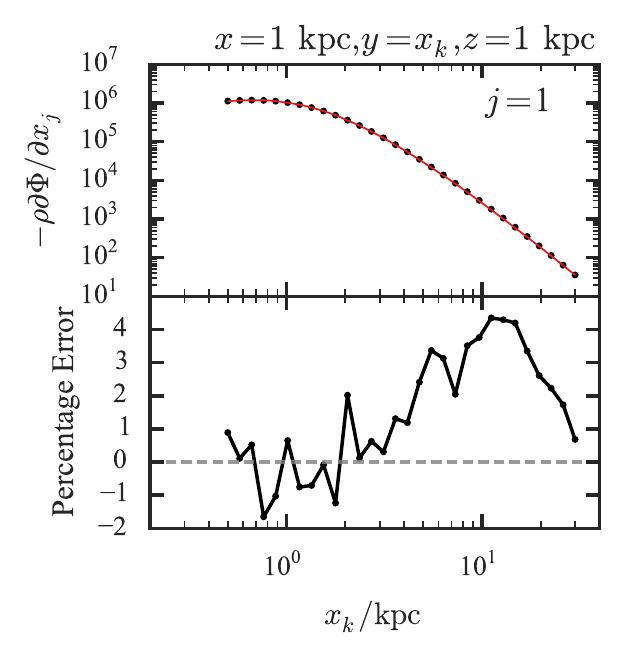}$$&
$$\includegraphics[bb = 7 8 171 181, width=0.31\textwidth]{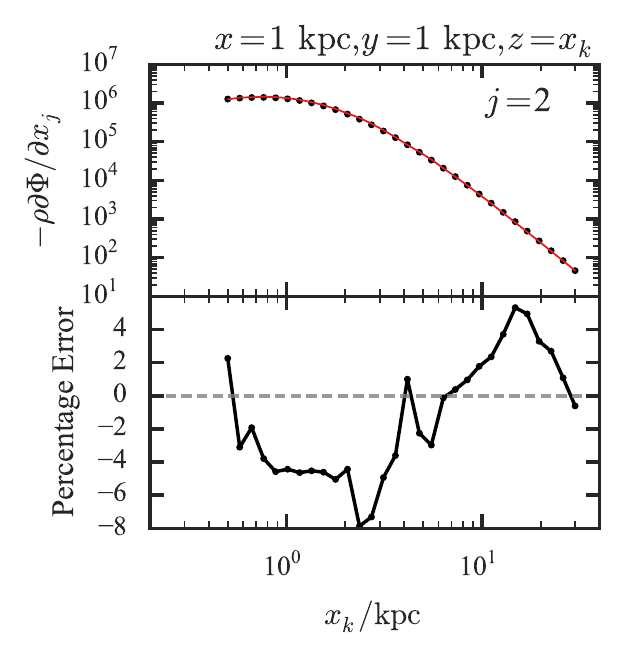}$$\\
$$\includegraphics[bb = 7 8 171 190, width=0.31\textwidth]{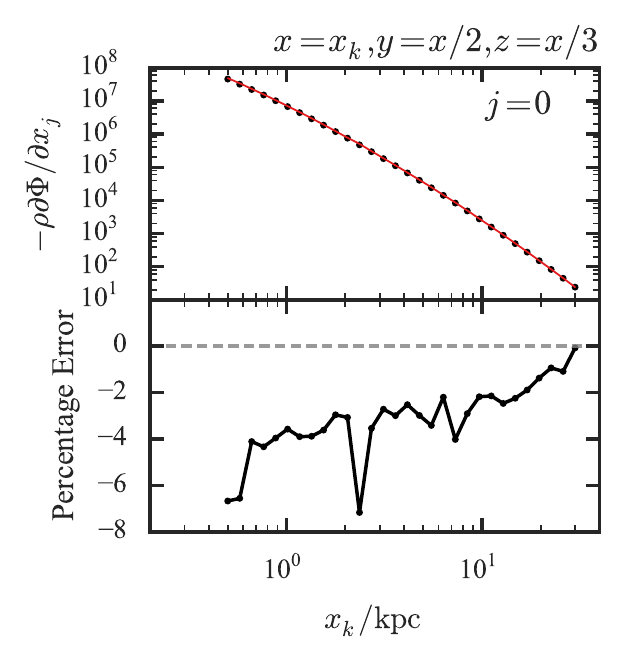}$$&
$$\includegraphics[bb = 7 8 171 182, width=0.31\textwidth]{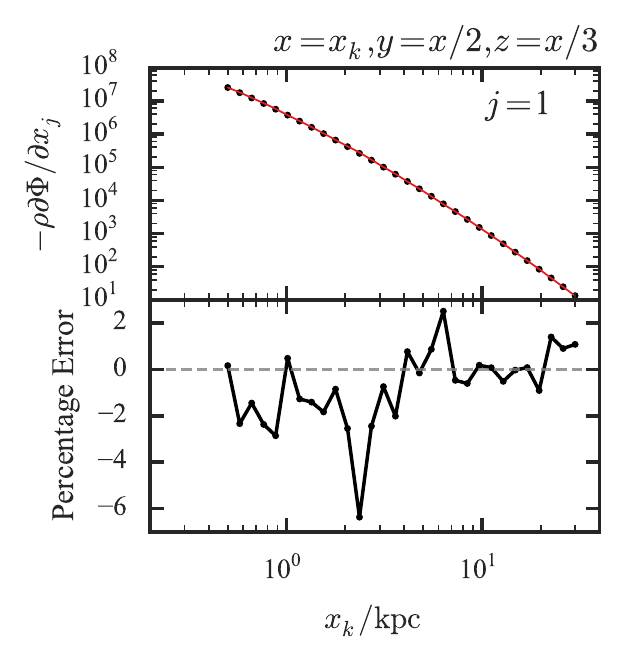}$$&
$$\includegraphics[bb = 7 8 171 182, width=0.31\textwidth]{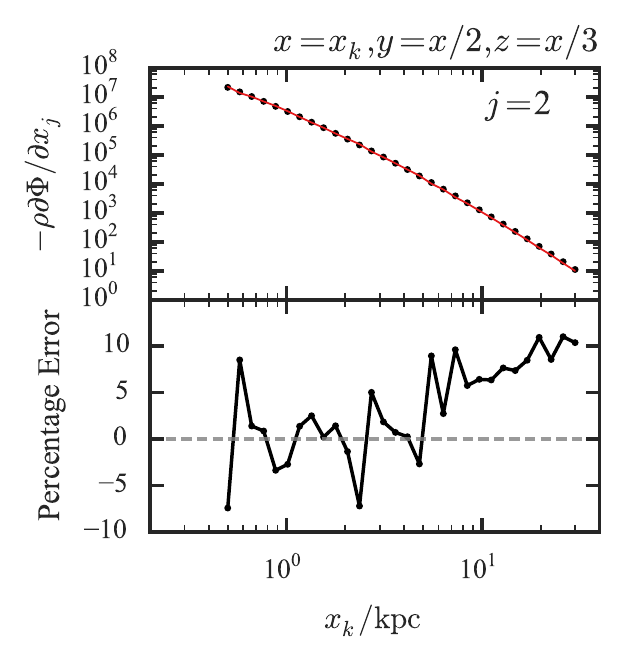}$$
\end{tabular}
\caption{Accuracy of Jeans' equation calculation for the radially-biased model ($\zeta=3.28$). In the top half of each panel we show $-\rho\,\upartial\Phi/\upartial x_j$ as a series of black dots and $\upartial(\rho\sigma^2_{ij})/\upartial x_i$ as a red line. In the bottom half we show the percentage error difference between these quantities. Each panel shows a single component, i.e. a single $j$, along the line parametrized by the coordinate $x_k$ and given above the top-right corner of each panel. The bottom three panels all correspond to the same line. The grey dashed line is the zero error line.}
\label{RadialJeans}
\end{figure*}

\begin{figure*}
\begin{tabular}{lll}
$$\includegraphics[bb = 7 8 171 190,width=0.31\textwidth]{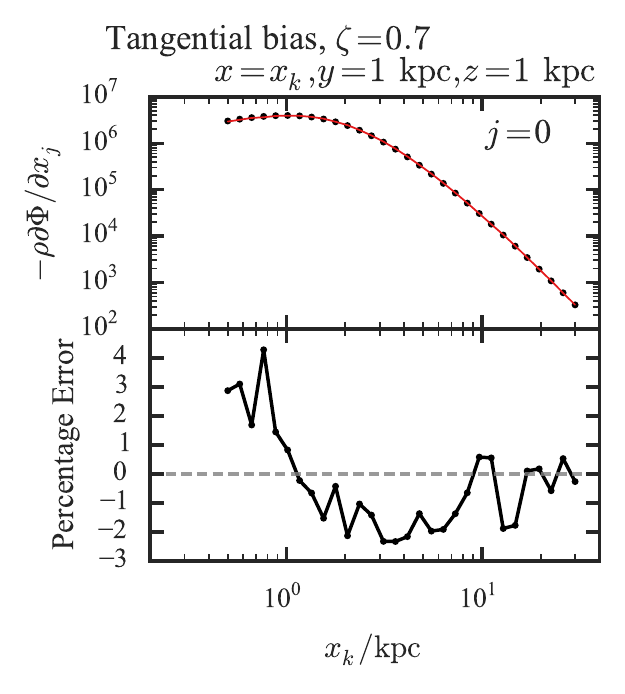}$$&
$$\includegraphics[bb = 7 8 171 181,width=0.31\textwidth]{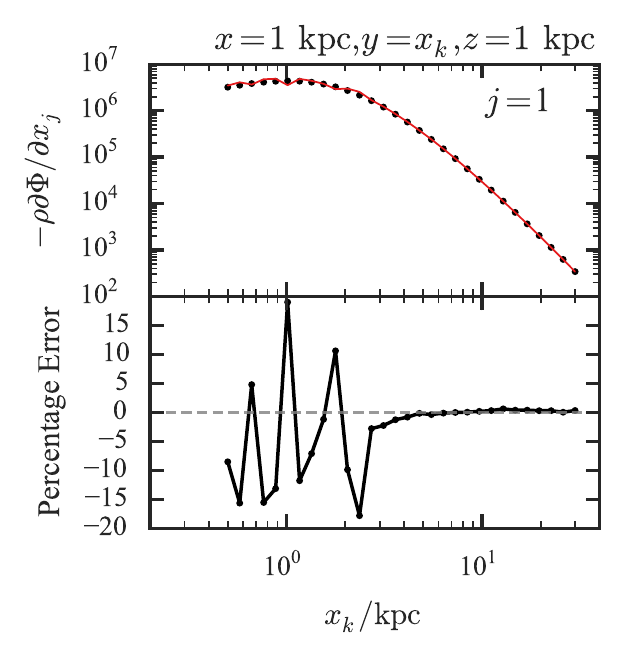}$$&
$$\includegraphics[bb = 7 8 171 181,width=0.31\textwidth]{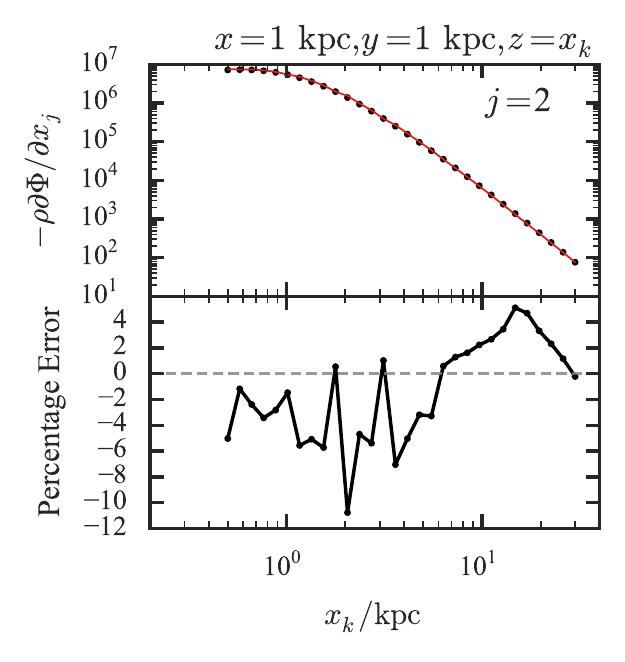}$$\\
$$\includegraphics[bb = 7 8 171 190,width=0.31\textwidth]{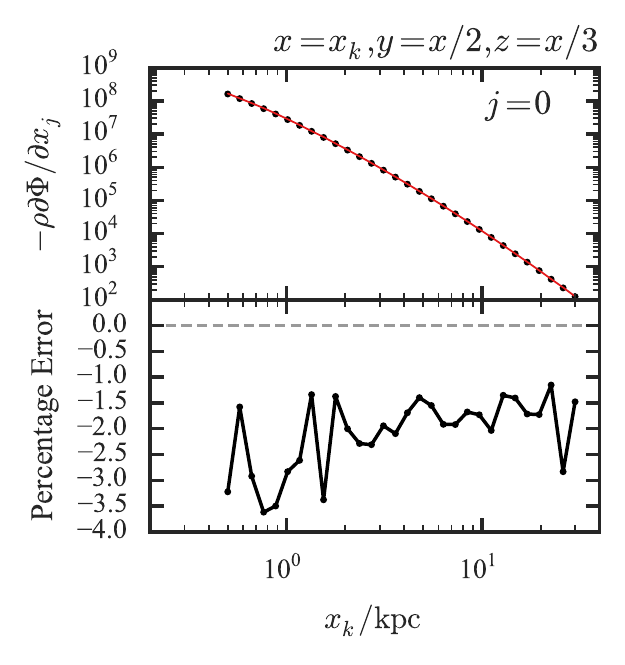}$$&
$$\includegraphics[bb = 7 8 171 182,width=0.31\textwidth]{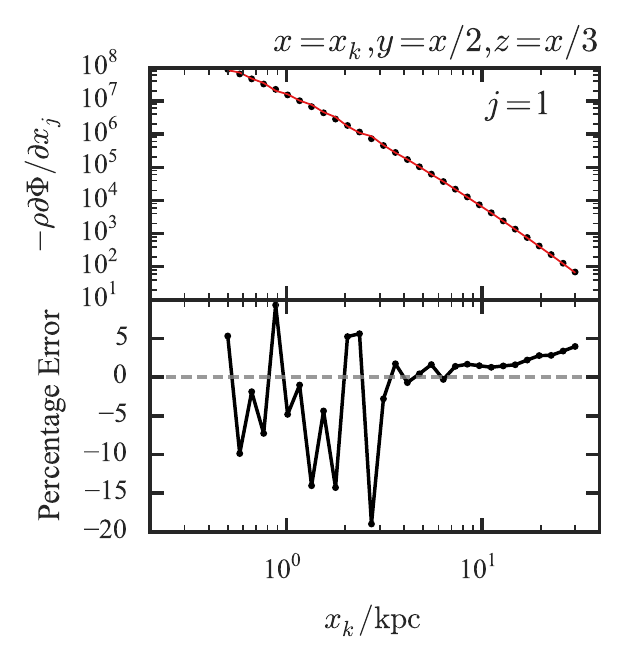}$$&
$$\includegraphics[bb = 7 8 171 182,width=0.31\textwidth]{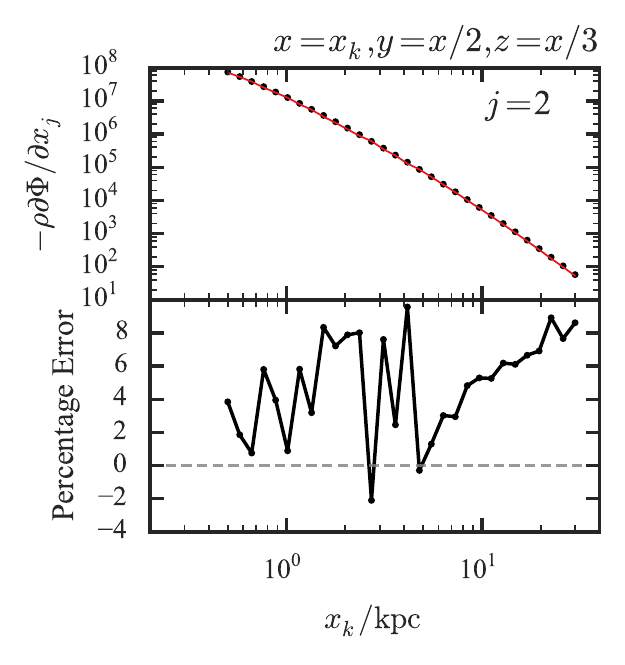}$$
\end{tabular}
\caption{Accuracy of Jeans' equation calculation for the tangentially-biased model ($\zeta=0.7$). In the top half of each panel we show $-\rho\,\upartial\Phi/\upartial x_j$ as a series of black dots and $\upartial(\rho\sigma^2_{ij})/\upartial x_i$ as a red line. In the bottom half we show the percentage error difference between these quantities. Each panel shows a single component, i.e. a single $j$, along the line parametrized by the coordinate $x_k$ and given above the top-right corner of each panele. The bottom three panels all correspond to the same line. The grey dashed line is the zero error line.}
\label{TangentialJeans}
\end{figure*}

A simple test of our action-based distribution functions is checking whether they satisfy these equations. The right hand side is calculated from analytic differentiation of the potential and multiplying by the density. The left hand side is found by numerically differentiating the three-dimensional integrals $\rho\sigma^2_{ij}$ and summing the appropriate contributions. Numerical differentiation of an integral leads to significant noise. To combat this we use an adaptive vectorised integration-rule cubature scheme implemented in the {\sc cubature} package from Steven Johnson ({http://ab-initio.mit.edu/wiki/index.php/Cubature}). Using a fixed-rule adaptive routine means the noise in the integrals is controlled such that the numerical derivatives are less noisy. In Figures~\ref{RadialJeans} and~\ref{TangentialJeans} we show how accurately the Jeans equations are satisfied along several lines through the potential for our two models. We plot each side of each Jeans equation for a choice of $j$ along a range of lines, along with the percentage error difference between the two sides of the equation. We avoid calculating the derivatives of the moments along the axes as the numerical differentiation is awkward there. In general, we find $\lesssim10\percent$ error for nearly all tested points with the majority having $\lesssim4\percent$ over a range of $\sim 8$ orders of magnitude.

Despite the large errors introduced by the action estimation scheme, we have produced a distribution function that satisfies the Jeans equations to reasonable accuracy. Even for the heavily radially-biased model, which has large contributions from the box orbits, the Jeans equations are well satisfied. This gives us confidence that models based on triaxial distribution functions can be constructed using the scheme we have presented.

\subsection{Error in the moments}

When comparing smooth models to data, we are primarily interested in whether
the action-estimation scheme produces accurate enough actions to reproduce
the features of the data. If the broad features of the data are on scales
larger than the errors in the individual actions, our method should be
sufficiently accurate to recover these features from an appropriate $f(\vJ)$.
In this case, the error in the moments is more important than the error in
the individual actions. Larger errors in the actions are expected to lead to
larger errors in the moments but the relationship between the two is unclear.
We have seen that, despite the presented method introducing large errors in
some actions, the moments of the \df{} are well recovered such that the
normalization is accurate to $0.6\percent$ and the Jeans' equations are
accurate to $\lesssim4\percent$. In this section, we develop some
understanding of how errors in the actions translate into errors in the
moments. Initially, we can make some progress by considering the
normalization of the \df{}: $\int\d^3\vx\,\d^3\vv\,f(\vx,\vv) =
\int\d^3\vJ\,\d^3\btheta\,f(\vJ)$.

Suppose we have a set of angle-action variables $(\vJ',\btheta')$ that are not the true angle-action variables $(\vJ,\btheta)$, we can relate the two sets via the generating function, $\mathcal{S}(\vJ,\btheta')$, such that
\begin{equation}
\begin{split}
\mathcal{S}(\vJ,\btheta') &= \vJ\cdot\btheta'+\sum_{\vn}\mathcal{S}_\vn(\vJ)\sin\vn\cdot\btheta',\\
\vJ' = \partial_{\btheta'}\mathcal{S} &= \vJ+\sum_{\vn}\vn\mathcal{S}_\vn(\vJ)\cos\vn\cdot\btheta',\\
\btheta = \partial_{\vJ}\mathcal{S} &= \btheta'+\sum_\vn\partial_\vJ\mathcal{S}_\vn(\vJ)\sin\vn\cdot\btheta'.
\end{split}
\label{Generating_Function}
\end{equation}
Now suppose we evaluate the normalization using $\vJ'(\vx,\vv)$. We transform
the volume element $\d^3\vJ\,\d^3\btheta$ to $\d^3\vJ\,\d^3\btheta'$ via the
Jacobian
\begin{equation}
\begin{split}
{\rm det}\Big(\frac{\partial \btheta}{\partial \btheta'}\Big)_\vJ
&= {\rm det}\Big(\mat{I}+\sum_\vn
\vn\otimes\partial_\vJ\mathcal{S}_\vn\cos\vn\cdot\btheta'\Big)\\
&\simeq |1+\sum_\vn\vn\cdot\partial_\vJ\mathcal{S}_\vn\cos\vn\cdot\btheta'|.
\end{split}
\label{dtdtprime_J}
\end{equation}
We assume that the approximate angle-action variables are sufficiently close to the true angle-action variables that the second term on the right-hand side is much less than unity.
Therefore, the difference in the normalization is given by
\begin{equation}
\begin{split}
\int\d^3\vJ\,&\d^3\btheta\,[f(\vJ)-f(\vJ')]=\int\d^3\vJ\,\d^3\btheta\,\partial_\vJ f \cdot(\vJ-\vJ')\\\simeq&\int\d^3\vJ\,\d^3\btheta'\,(1+\sum_\vn\vn\cdot\partial_\vJ\mathcal{S}_\vn\cos\vn\cdot\btheta')\\&\times\Big[-\sum_{\vm}(\vm\cdot\partial_\vJ f)\mathcal{S}_\vm(\vJ)\cos\vm\cdot\btheta'+\cdots\Big].
\end{split}
\end{equation}
We see that, after integrating over $\btheta'$, all terms with odd powers of trigonometric functions vanish, and the leading order error in the normalization is
\[
-4\pi^3\int\d^3\vJ\,\sum_\vn (\vn\cdot\partial_\vJ f) (\vn\cdot\partial_\vJ \mathcal{S}_\vn)\mathcal{S}_\vn.
\]
This term is second order in the Fourier components of the generating function. Note that the sign of this term is unclear as both the $\mathcal{S}_\vn$ and $\partial_\vJ \mathcal{S}_\vn$ can be positive and negative. We anticipate $\partial_\vJ f$ is negative such that the density falls with radius. From equation~\eqref{Generating_Function} we find that to leading order the errors in the angle-action variables averaged over an orbit are
\[
\Delta J_i\leq\sqrthalf\sum_\vn|\mathcal{S}_\vn n_i|$ {\rm\,and\,} $\Delta\theta_i\leq\sqrthalf\sum_\vn|\partial_{J_i} \mathcal{S}_\vn|.
\]
Therefore, we can see that the error in the normalization is approximately
first order in the error in the actions, $\Delta J_i$, the error in the
angles, $\Delta\theta_i$, and the gradient of the distribution function,
$\partial_{J_i}f$. Therefore, we anticipate that the relative error in the
normalization will be small when $\Delta\vJ\ll f(\partial_\vJ f)^{-1}$ for
all points in action space. This is essentially the expected result.
Consider the distribution functions of \cite{Binney2014}: these are of the
form $f(J_z)\sim\exp(-\nu J_z/\sigma_z^2)$ such that $\Delta J_z\ll
\sigma_z^2/\nu$ for a good estimate of the normalization, where $\nu$ is the
vertical epicycle frequency. Near the Sun $\nu\approx0.1{\rm Myr}^{-1}$ and
$\sigma_z\approx 30\kms$ so $\Delta J_z\ll10\kpc\kms$. For the distribution
function considered in the previous section we require $\Delta
J\ll\hbox{max}(J,J_0)$.

\begin{figure}
$$\includegraphics[bb = 7 9 322 479, width=\columnwidth]{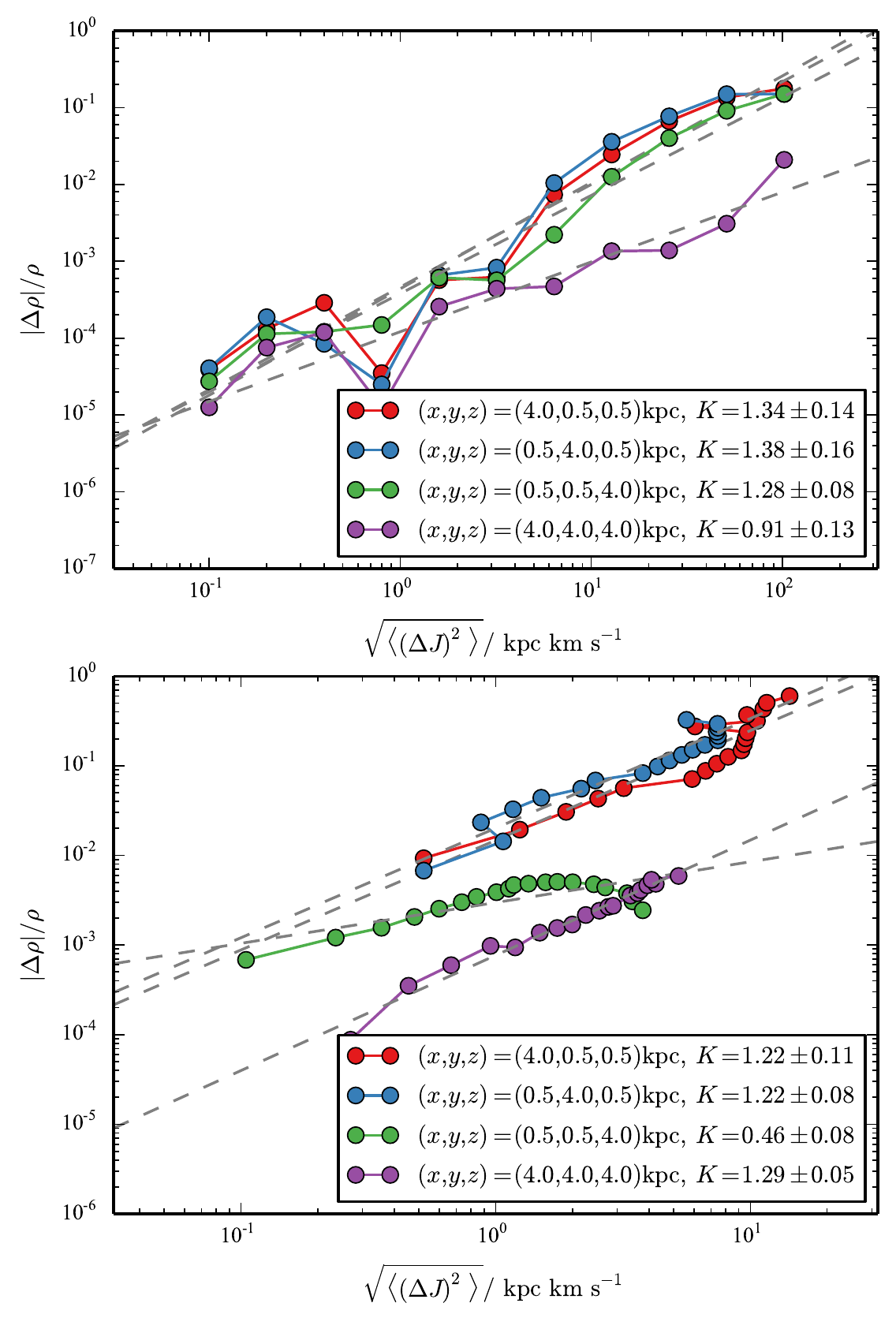}$$
\caption{Relative error in the density as a function of distribution-function-weighted RMS action error. The top panel shows the error when scattering by some fixed $\sqrt{\langle(\Delta J)^2\rangle}$ and the bottom panel shows the error when using the wrong $\alpha$ value. Each line shows the density at a fixed Cartesian position shown in the legend along with the gradients of the lines, $K$.}
\label{DensityError}
\end{figure}

For the moments of the distribution function we expect similar results but we
are not able to explicitly calculate the leading order errors in these
quantities. Instead we briefly show how the error in the density changes with
the error in the actions. We begin by calculating the density from a triaxial
\df{} for which we know the true density -- the perfect ellipsoid
\citep{deZeeuw1985}, which has density profile
\begin{equation}
\rho(x,y,z) = \frac{\rho_0}{(1+m^2)^2},
\end{equation}
where
\begin{equation}
m^2\equiv\frac{x^2}{x_P^2}+\frac{y^2}{y_P^2}+\frac{z^2}{z_P^2},\>x_P\geq y_P\geq z_P\geq 0.
\end{equation}
We set $\rho_0 = 7.2\times10^8 M_\odot{\kpc}^{-3},$ $x_P = 5.5\kpc,$ $y_P = 4.5\kpc$ and $z_P = 1\kpc$. The actions in this potential can be found exactly using the scheme presented above by setting $\alpha=-x_P^2$ and $\beta=-y_P^2$. We use the tangentially-biased \df{} of the previous section. For four different Cartesian positions we calculate the true density, and then proceed to calculate the density when the logarithm of the actions are scattered normally by some fixed amount $\sqrt{\langle(\Delta J)^2\rangle}$. We plot the error in the density as a function of $\Delta J$ in Fig.~\ref{DensityError}. We find that the relative error in the density goes as $\sim\Delta J^{1.3}$ for the three densities near the axes, and is flatter for the density at $(x,y,z)=(4,4,4)\kpc$.

This procedure is artificial as we have used non-canonical coordinates to
evaluate the density. Therefore, for a fuller test we instead choose to
calculate the density using the St\"ackel fudge scheme but changing $\alpha$
away from the truth. In this case, the error in the density is systematic. In
the lower panel of Fig.~\ref{DensityError} we plot the density error from
this procedure as a function of the distribution-function-weighted RMS action
errors. Again we find the density goes as $\sim\sqrt{\langle(\Delta
J)^2\rangle}^{1.2}$ for all but the density on the $z$ axis. As we are
changing $\alpha$ this does not significantly affect the actions of the
long-axis loop orbits, which dominate the density budget along the $z$ axis.

\section{Extending the scope of torus mapping}\label{sec::twoup}

We have seen that the triaxial St\"ackel fudge can produce large errors in
the actions for some orbits, but that the moments of an action-based \df{}
are nevertheless well recovered using the method. However, even if one
requires accurate actions, the St\"ackel fudge can be valuable because it
enables one to construct a torus through a given point $(\vx,\vv)$ by torus
mapping rather than orbit integration \citep{SandersBinney2014}. This option
may be essential because torus mapping works even in a chaotic portion of
phase space \citep{Kaasalainen1995}, while the approach based on orbit
integration is already problematic when resonantly trapped orbits take up a
significant portion of phase space, and it breaks down with the onset of
chaos.

We proceed as follows. First we use the St\"ackel fudge to obtain approximate
actions
\[
\vJ=\vJ_{\rm St}(\vx,\vv).
\]
 Then by torus mapping we obtain the torus with actions $\vJ$. On
account of errors, the given point $(\vx,\vv)$ will not lie on the
constructed torus, but one can identify the nearest point $(\vx',\vv')$ that
{\it does} lie on the torus. We find this point by minimising the tolerance
\begin{equation}
\eta = |\bs{\Omega}|^2|\vx(\btheta)-\vx|^2+|\vv(\btheta)-\vv|^2,
\end{equation}
with respect to the angle, $\btheta$, on the torus. $\bs{\Omega}$ is the
frequency vector of the constructed torus. We use the St\"ackel fudge estimate of the angles as an initial guess for the minimisation. For the point $(\vx',\vv')$ we know the true actions:
\[
\vJ=\vJ_{\rm t}(\vx',\vv').
\]
 Now we use the St\"ackel fudge to obtain  approximate actions for this point
\[
\vJ'=\vJ_{\rm St}(\vx',\vv').
\]
 If, as we expect, the errors in the St\"ackel fudge are
systematic rather than random, then
\[
\vJ_{\rm t}(\vx,\vv)=\vJ_{\rm St}(\vx,\vv)+\vDelta
\]
 with $\vDelta$ a slowly varying function of phase-space position.
So a better estimate of the true actions of the
original point $(\vx,\vv)$ is
 \[
\vJ''=\vJ+\vDelta=\vJ+(\vJ-\vJ')=2\vJ-\vJ'.
\]
 If one is of a nervous disposition, one now uses torus mapping to construct
the torus with actions $\vJ''$ and seeks the point on this torus that is
closest to the given point and applies the St\"ackel fudge there, and so on.
This cycle can be repeated until the nearest point on the constructed torus
satisfies some tolerance $\eta=\eta_*$.

\begin{figure}
$$\includegraphics[bb = 7 9 278 323, width=\columnwidth]{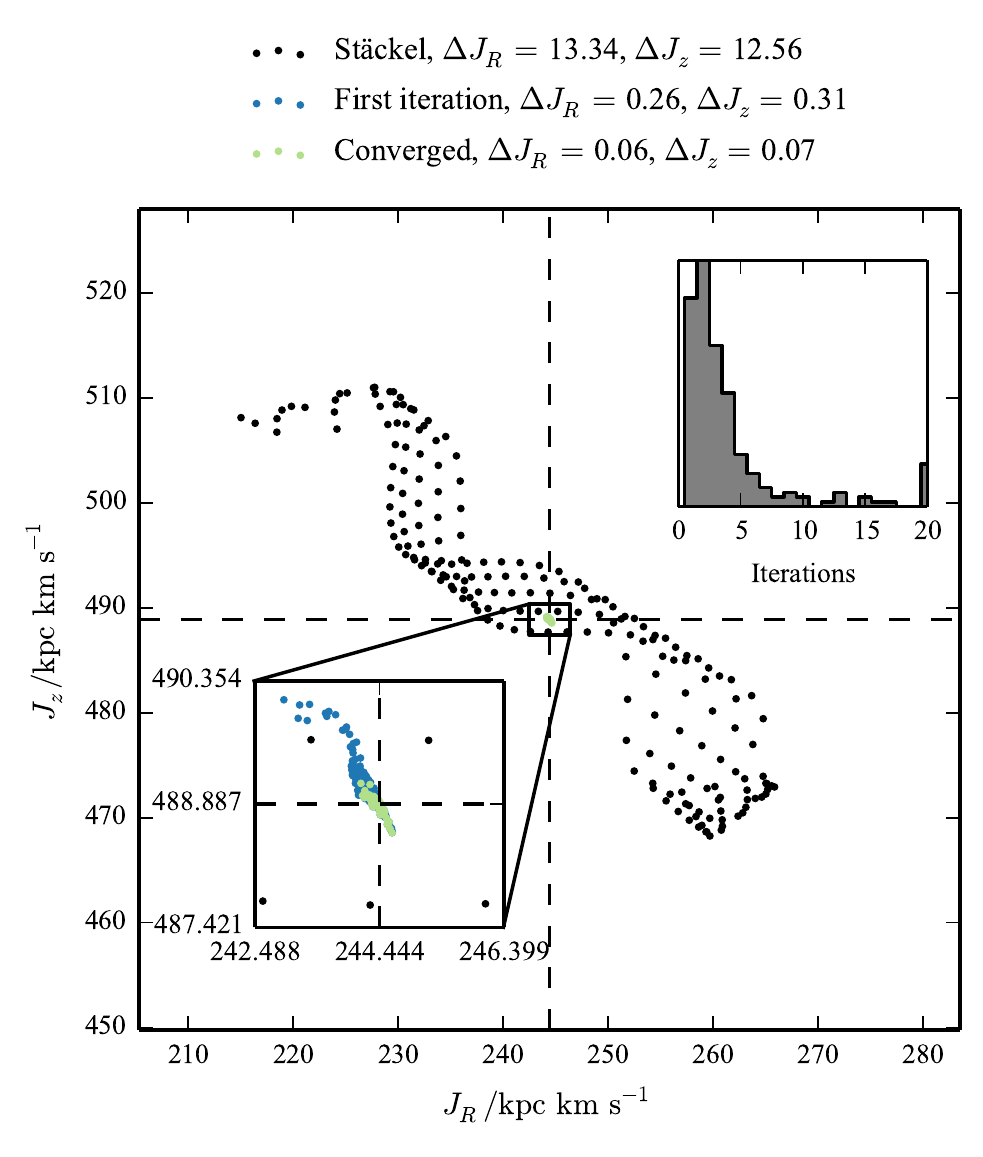}$$
\caption{Illustration of the iterative torus scheme presented in Section~\ref{sec::twoup}. The black points show the initial estimate of the actions for a series of points along an orbit calculated using the axisymmetric St\"ackel fudge of \protect\cite{Binney2012}. The dark blue points show the improved estimate from a single iteration of the torus scheme, and the light green points show the estimate after the scheme was deemed to converge to an accuracy of $\eta_*=(0.1\kms)^2$. The top-right inset shows a histogram of the number of iterations required to reach this accuracy. Above the plot we show the standard deviation of the action estimates for each set of points. The black dashed lines give the true action estimate. The bottom-left inset shows a zoom-in of the central region of the main plot.}
\label{IterativeTorus}
\end{figure}

In Fig.~\ref{IterativeTorus} we show an illustration of this procedure for the axisymmetric case. We use the axisymmetric St\"ackel fudge as given in Section~\ref{Sec::axisym} and the torus construction code as presented in \cite{McMillanBinney2008}. For the axisymmetric St\"ackel fudge we set $\gamma=-1\kpc^2$ and $\alpha=-20\kpc^2$, such that the foci are at $z=
\pm\sqrt{\gamma-\alpha}\approx\pm4.4\kpc$. We construct a torus of actions $(J_r,L_z,J_z) = (244.444,3422.213,488.887)\kpc\kms$ in the ``best'' potential from \cite{McMillan2011}. This potential is an axisymmetric multi-component Galactic potential consisting of two exponential discs representing the thin and thick discs, an axisymmetric bulge model from \cite{BissantzGerhard} and an NFW dark halo. The parameters of the mass model were chosen to satisfy recent observational constraints. We produce a series of $(\vx,\vv)$ points on the constructed torus. The axisymmetric St\"ackel fudge gives errors of $\Delta J_i = 13\kpc\kms$ in the recovery of the actions for these points. After one iteration of the above procedure we find actions accurate to $\Delta J_i \approx 0.3\kpc\kms$, and after the procedure has converged to a tolerance $\eta_*=(0.1\kms)^2$ the actions are accurate to $\Delta J_i \approx 0.07\kpc\kms$. The majority of $(\vx,\vv)$ points converge in less than five iterations. We limited the number of iterations to $20$ and a few $(\vx,\vv)$ did not converge within $20$ iterations. This is due to non-linear behaviour of both the torus construction and St\"ackel code in small phase-space volumes. It is clear, however, that only one iteration is required for a substantial improvement in the actions, and it is hard to make a case for more iterations.

\begin{figure}
$$\includegraphics[bb = 7 8 235 330, width=\columnwidth]{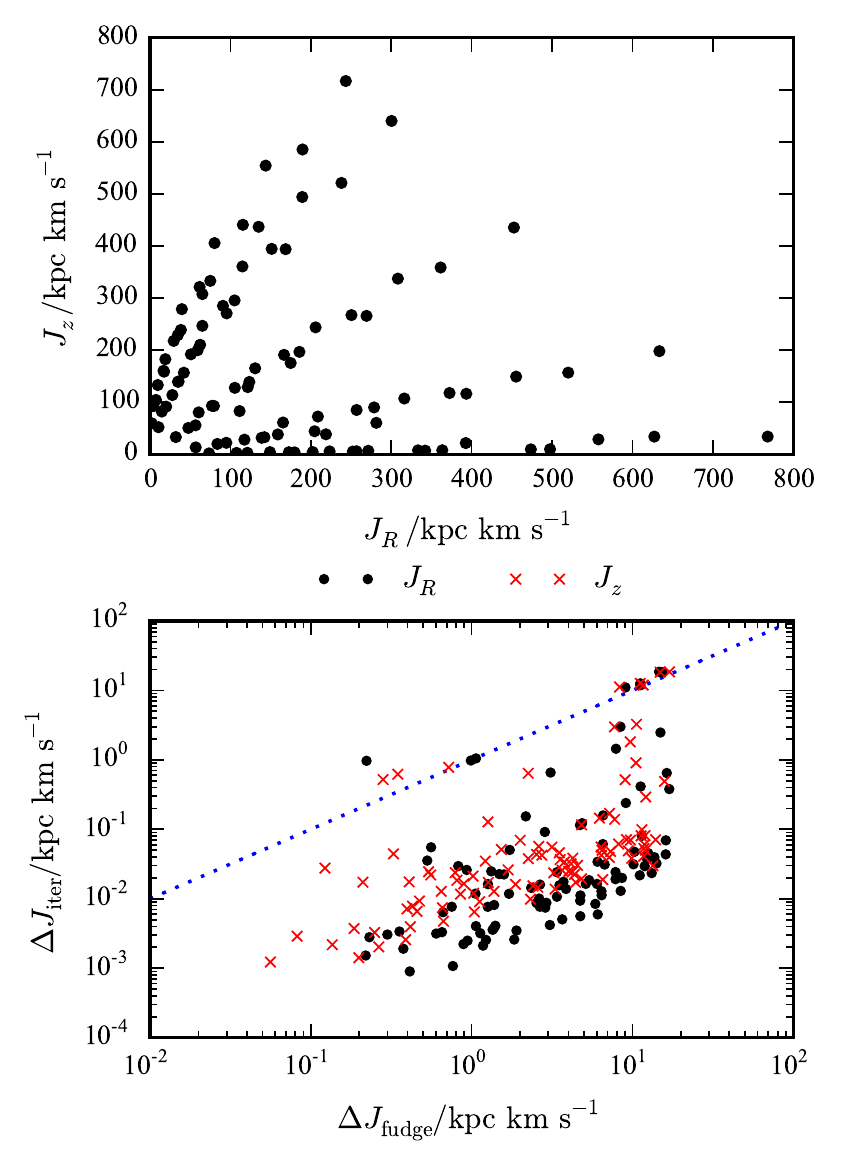}$$
\caption{Actions of a selection of orbits shown in the top panel, along with the absolute error in the action for these orbits from the axisymmetric St\"ackel fudge of \protect\cite{Binney2012} plotted against the absolute error from the iterative torus procedure for a range of orbits in the bottom panel. The black dots show the results for the radial action, whilst the red crosses show the results for the vertical action. The blue dotted line is the 1:1 line.}
\label{IterativeTorus_Lots}
\end{figure}

Finally, we demonstrate how the above method operates for a range of
high-action orbits. We integrate a series of orbits in the ``best'' potential
from \cite{McMillan2011} launched at $5$ linearly-spaced points $x_0$ along
the $x$-axis such that $4\kpc\leq x_0\leq 12\kpc$. We launch the orbits with
velocity $\bs{v}=(v_1\cos\theta,v_0,v_1\sin\theta)$ where
$v_0=\sqrt{x_0\partial_x\Phi(x_0,0,0)}$ and we choose $4$ linearly-spaced
values of $v_1$ such that $0.5v_0\leq v_1 \leq0.8v_0$ and $5$ linearly-spaced
angles $\theta$ such that $0.2\rad\leq\theta\leq \half\pi\rad$. The range of
radial and vertical actions for this collection of orbits is shown in the top
panel of Fig.~\ref{IterativeTorus_Lots} and is approximately
$1\kpc\kms\lesssim J_r \lesssim 800 \kpc\kms$ and $1\kpc\kms \lesssim J_z
\lesssim 800\kpc\kms$ (these are calculated as the mean of the fudge
estimates along the orbit). For each orbit, we find the standard deviation of
the action estimates from the axisymmetric St\"ackel fudge method and the
iterative torus method for $10$ widely time-separated phase-space points
along the orbit. We use $\eta_*=(0.1\kms)^2$, limit the maximum number of
iterations to $5$ and construct the tori with a relative error of
$\sim1\times10^{-4}$. We plot the results in Fig.~\ref{IterativeTorus_Lots}.
We see the majority of orbits have lower iterative torus errors than fudge
errors and follow a broad line that lies approximately two to three orders of
magnitude beneath the 1:1 line. However, there are several orbits that lie
close to the 1:1 line indicating the procedure has not converged to a greater
accuracy than the initial accuracy produced by the St\"ackel fudge. These
orbits are either near-resonant so require more careful action assignment
\citep{Kaasalainen1995}, or have one action very much greater than the other
(near radial or shell orbits) so require more accurate torus construction
than our automated procedure has allowed. It is clear from the plot that none
of the iterative procedures have diverged significantly as all points lie
near or well below the 1:1 line.

The results for lower-action disc-type
orbits are superior to those presented here. However, by and large the
St\"ackel fudge action estimates for these orbits are sufficiently accurate
for much scientific work \citep{Piffl2014} so the iterative procedure is
probably not required. One realistic application of the presented method is
the modelling of tidal streams: \cite{Sanders2014} used the expected angle and frequency structure of a stream in the correct potential to constrain the potential from a stream simulation. The frequencies of the collection of orbits explored here range from $10$ to $120\kpc^{-1}\kms$. For the majority of orbits the error in the frequencies recovered from the fudge are $\lesssim10\percent$ with the majority having errors of a few per cent, whilst the iterative torus approach reduces the errors to approximately $0.01\percent$ for all orbits apart from those with large action errors discussed previously. The $10^4M_\odot$ stream used in \cite{Sanders2014} had a frequency width to absolute frequency ratio of $\sim0.1\percent$ so the iterative torus approach seems well suited to modelling of streams. In conclusion, we have demonstrated that the presented algorithm has the capability to produce accurate actions for a wide range of orbits.

\section{Conclusions}\label{Sec::TriaxConclusions} We have presented a method
for estimating the actions in a general triaxial potential using a St\"ackel
approximation. The method is an extension of the St\"ackel fudge introduced
by \cite{Binney2012} for the axisymmetric case. We have investigated the
accuracy of the method for a range of orbits in an astrophysically-relevant
triaxial potential. We have seen that the recovery of the actions is poorest
for the box orbits, which probe a large radial range of the potential, and
much better for the loop orbits, which are confined to a more limited radial
range. The only parameters in the method are the choice of the focal
positions $\Delta_i$, which are selected for each input phase-space point. We
have detailed a procedure for selecting these based on the energy of the
input phase-space point. This choice is not optimal but, by adjusting
$\Delta_i$, we can, at best, increase the accuracy of the actions of a factor
of two for the triaxial NFW potential considered. However, to achieve this
accuracy requires additional computation for each input phase-space point
(e.g. orbit integration). For general potentials the best action estimates
will be achieved when locally (over the region a given orbit probes) the
potential is well approximated by some St\"ackel potential. Many potentials
of interest are not well fitted \emph{globally} by St\"ackel potentials so
the accuracy of the action estimates will deteriorate for orbits with large
radial actions.

The advantage of this method over other methods for estimating the actions in
a triaxial potential is speed. Unlike the convergent method introduced by
\cite{SandersBinney2014}, we obtain the actions without integrating an orbit
-- we only use the initial phase-space point.  We have only to evaluate
several algebraic expressions, find the limits of the orbits in the $\tau$
coordinate and perform Gaussian quadrature.  These are all fast calculations.
However, this speed comes at the expense of sometimes disappointing accuracy.
If accurate results are required, the St\"ackel fudge can be combined with
torus mapping to form a rapidly convergent scheme for the determination of
$\vJ(\vx,\vv)$. We demonstrated how such a scheme performed in the
axisymmetric case and found a single torus construction provided a high level
of accuracy that is not significantly improved by further torus
constructions.

We went on to construct, for the first time, triaxial stellar systems from a
specified \df{}s $f(\vJ)$ in Section~\ref{Sec::DF}. We demonstrated the mass
of these models is well recovered using the St\"ackel fudge, and we showed
how the error in the density of these models varies as a function of the
action error. Notwithstanding the errors in individual actions, both a
radially-biased model and a tangentially-biased model satisfy the Jeans
equations to good accuracy. This is because individual errors largely cancel
during integration over velocities when computing moments such as the
density $\rho(\vx)$ and the pressure tensor $\rho\sigma_{ij}^2(\vx)$.

The results presented in this paper have focussed on a limited range of
astrophysically-relevant models: we have used a single specific NFW potential
and two simple distribution functions that depend on a linear sum of the
actions. However, we anticipate that our results will extend to more general
distribution functions. We have investigated analytically how the
normalization of the distribution function varies with the error in the
action estimates and shown that the normalization is well recovered provided
the error in the actions is smaller than the action scale over which the
distribution function varies significantly i.e. $\Delta\vJ\ll f(\partial_\vJ
f)^{-1}$. Therefore, the recovery of the moments is expected to be most
accurate for distribution functions with shallow radial density profiles and
to
deteriorate with the steepness of the required profile.

Whilst the scheme presented here does not give accurate enough actions for
working with streams \citep{Sanders2014} we have shown that it is an
appropriate and powerful tool for constructing models from specified \df{}s
$f(\vJ)$. A key property of \df{}s of the form $f(\vJ)$ is that they can be
trivially added to build up a multi-component system. Hence the ability to
extract observables from \df{}s of the form $f(\vJ)$ is likely to prove
extremely useful for interpreting data on both external galaxies
\citep{Atlas3D} and our Galaxy, in which components such as the stellar and
dark haloes may be triaxial, and the bulge certainly is.

\section*{Acknowledgments}
We thank Paul McMillan for provision of the torus construction machinery used in Section~\ref{sec::twoup} and the Oxford Galactic Dynamics group for helpful comments. JLS acknowledges the support of STFC. JB was
supported by STFC by grants R22138/GA001 and ST/K00106X/1. The research
leading to these results has received funding from the European Research
Council under the European Union's Seventh Framework Programme
(FP7/2007-2013) / ERC grant agreement no.\ 321067.

{\footnotesize{
\bibliographystyle{mn2e-2}
\bibliography{triaxial_stackel}
}}

\appendix
\section{Angles and frequencies}\label{Appendix::Angles}
With the framework presented in Section~\ref{Sec::Triaxapproximation} we are also in a position to find the angles, $\btheta$, and frequencies, $\bs{\Omega}$. Following \cite{deZeeuw1985} we write
\begin{equation}
\begin{split}
\frac{\partial E}{\partial E} &= 1 = \sum_{\tau=\lambda,\mu,\nu}\Omega_{\tau}\frac{\partial J_\tau}{\partial E},\\
\frac{\partial E}{\partial a} &= 0 = \sum_{\tau=\lambda,\mu,\nu}\Omega_{\tau}\frac{\partial J_\tau}{\partial a},\\
\frac{\partial E}{\partial b} &= 0 = \sum_{\tau=\lambda,\mu,\nu}\Omega_{\tau}\frac{\partial J_\tau}{\partial b}.
\end{split}
\end{equation}
Inversion of these equations gives, for instance,
\begin{equation}
\Omega_\lambda = \frac{1}{\Gamma}\frac{\partial(J_\mu,J_\nu)}{\partial(a,b)} \text{ where }\Gamma = \frac{\partial(J_\lambda,J_\mu,J_\nu)}{\partial(E,a,b)},
\end{equation}
and $\Omega_\mu$ and $\Omega_\nu$ are given by cyclic permutation of $\{\lambda,\mu,\nu\}$. To find the derivatives of $J_\tau$ with respect to the integrals we differentiate equation~\eqref{Eq::action} under the integral sign at constant $\tau$. From equation~\eqref{Eq::EqnOfMotion_JK} we know $p_\tau(\tau,E,A_\tau,B_\tau)$. We note that
\begin{equation}
\begin{split}
\frac{\partial}{\partial a}\Big|_\tau &= \frac{\partial A_\tau}{\partial a}\Big|_\tau\frac{\partial}{\partial A_\tau}\Big|_\tau = \frac{\partial}{\partial A_\tau}\Big|_\tau,\\
\frac{\partial}{\partial b}\Big|_\tau &= \frac{\partial B_\tau}{\partial b}\Big|_\tau\frac{\partial}{\partial B_\tau}\Big|_\tau = \frac{\partial}{\partial B_\tau}\Big|_\tau,
\end{split}
\end{equation}
as $A_\tau=a-C_\tau$ and $B_\tau=b+D_\tau$ where $C_\tau$ and $D_\tau$ are independent of $\tau$.
The required derivatives are
\begin{equation}
\begin{split}
\frac{\partial p_\tau}{\partial E}\Big|_\tau &= \frac{\tau^2}{4p_\tau(\tau+\alpha)(\tau+\beta)(\tau+\gamma)},\\
\frac{\partial p_\tau}{\partial a}\Big|_\tau &= -\frac{1}{\tau}\frac{\partial p_\tau}{\partial E}\Big|_\tau,\\
\frac{\partial p_\tau}{\partial b}\Big|_\tau &= \frac{1}{\tau^2}\frac{\partial p_\tau}{\partial E}\Big|_\tau.
\end{split}
\end{equation}
Note that $p_\tau$ can vanish at the limits of integration. The change of variables
\[
\tau = \hat{\tau}\sin\vartheta+\bar{\tau};\>\bar{\tau} = \frac{1}{2}(\tau_-+\tau_+);\>\hat{\tau} = \frac{1}{2}(\tau_+-\tau_-)
\]
causes the integrand to go smoothly to zero at the limits.
To find the angles, we use the generating function, $W(\lambda,\mu,\nu,J_\lambda,J_\mu,J_\nu)$, given by
\begin{equation}
W = \sum_{\tau=\lambda,\mu,\nu} W_\tau = \sum_{\tau=\lambda,\mu,\nu} \int_{\tau_-}^\tau \d\tau'p_\tau' + \mathcal{F}_\tau(p_\tau,\vx) \int_{\tau_-}^{\tau_+} \d\tau'|p_\tau'|.
\end{equation}
$\mathcal{F}_\tau$ are factors included to remove the degeneracy in the $\tau$ coordinates such that $\theta_\tau$ covers the full range $0$ to $2\pi$ over one oscillation in the Cartesian coordinates. These factors can be written in the form
\begin{equation}
\begin{split}
\mathcal{F}_\lambda (p_\lambda,\vx) &= \Pi(\lambda_-+\alpha)\Theta(-x)+\Theta(-p_\lambda),\\
\mathcal{F}_\mu (p_\mu,\vx) &=  \Pi(\mu_-+\beta)[\Theta(-y)+\Theta(-p_\mu)]\\
							&	+\Pi(\nu_++\beta)\Pi(\mu_++\alpha)[\half+\Theta(-x)]\\
							&	+\Pi(\nu_++\beta)\Pi(\lambda_-+\alpha)\Theta(-p_\mu),\\
\mathcal{F}_\nu (p_\nu,\vx) &= \Theta(-z)+\Theta(-p_\nu),
\end{split}
\end{equation}
where $\Theta$ is the Heaviside step function and $\Pi$ is one when its argument is zero and zero otherwise. The $\Pi$-function in $\mathcal{F}_\lambda$ takes care of the cases when the orbit is a box or inner long-axis loop. The $\Pi$-functions in $\mathcal{F}_\mu$ take care of the cases when the orbit is a short-axis loop or a box, an outer long-axis loop, and an inner long-axis loop respectively. The angles are given by
\begin{equation}
\theta_\tau = \frac{\partial W}{\partial J_\tau} = \sum_{I=E,a,b}\frac{\partial W}{\partial I}\frac{\partial I}{\partial J_\tau}.
\end{equation}
The first term on the right is, up to factors, the indefinite integral of the derivatives of $J_\tau$ with respect to the integrals found previously, whilst the second term is found from inverting these derivatives. We have chosen the zero-point of $\theta_\tau$ to correspond to $\tau=\tau_-$, $p_\tau>0$ and $\dot{x}_i\geq0$ for all $i$, except for the outer long-axis loop orbits which have $\theta_\mu=0$ at $\mu=-\alpha$, $p_\tau>0$ and $\dot{x}_i\geq0$. Note that the angles are the $2\pi$ modulus of the $\theta_\tau$ found from the above scheme.

\begin{figure}
$$\includegraphics[bb = 8 9 309 478, width=\columnwidth]{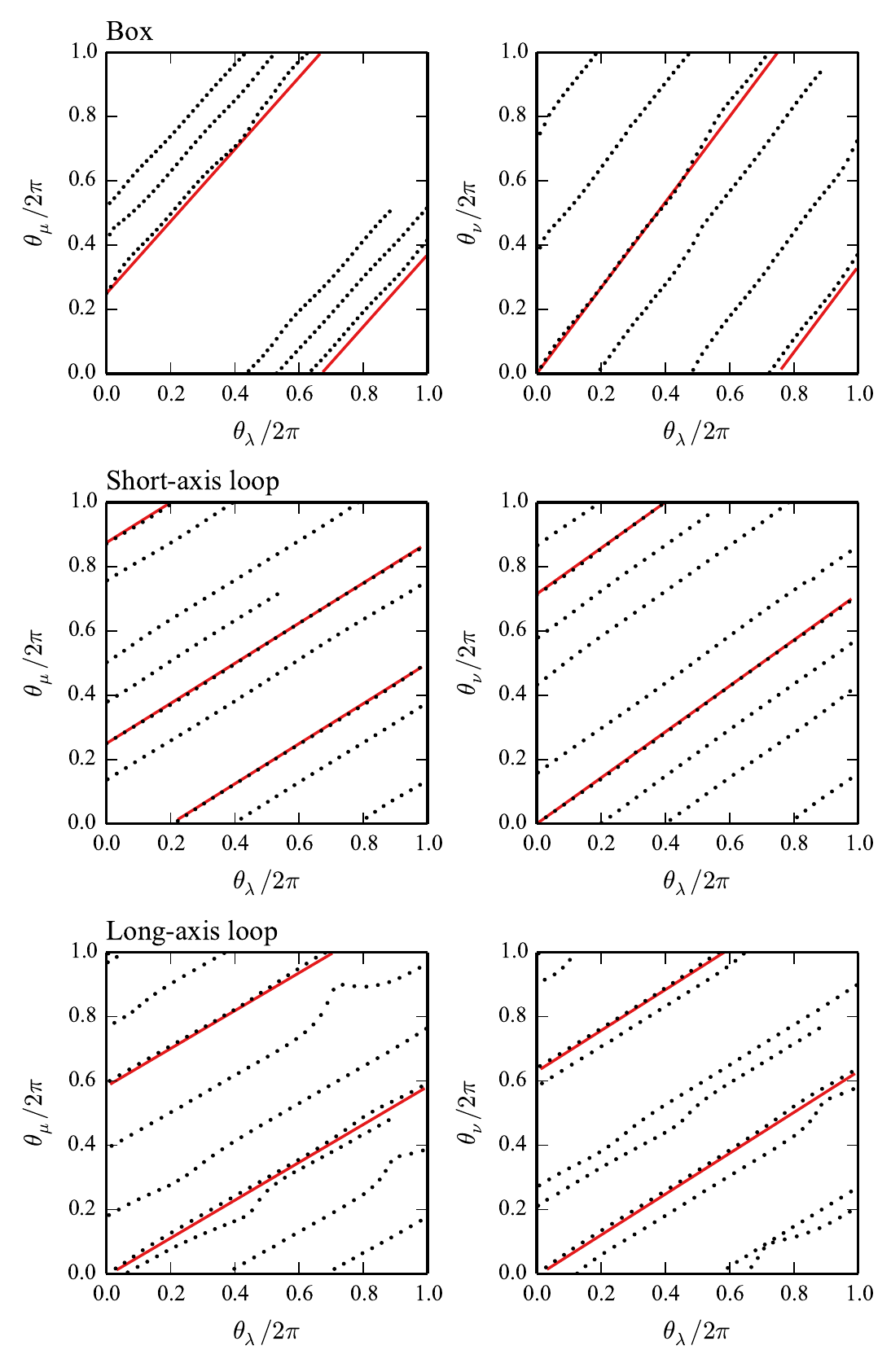}$$
\caption{Angles calculated using the triaxial St\"ackel fudge presented in this paper for three different orbits in the triaxial NFW potential. The solid red lines show the angles calculated from the initial angle estimate and the frequency estimates for approximately one period.}
\label{Angles}
\end{figure}

In Fig.~\ref{Angles} we show the angles calculated from the St\"ackel fudge for the three orbits investigated in Section~\ref{Sec::Accuracy}. We use the automatic choice of $\Delta_i$ for the box and short-axis loop orbit, and the choice that minimises the spread in actions for the long-axis loop orbit. The short-axis loop orbit shows the expected straight-line structure in the angle coordinates, whilst for the long-axis loop and box orbits there is clear deviation from this expected straight line. We also show the angles calculated using the initial angle estimate and the average of the frequency estimates along the orbit. We see that they are well recovered but after approximately one period the error in the frequencies is sufficient for these angles to deviate from the angle estimates.

The standard deviations in the frequencies are reasonably large. For the box orbit, the mean frequencies are given by $\vOmega = (18.1,20.3,24.3)\kpc^{-1}\kms$ with errors $\Delta\vOmega=(0.2,0.8,1.3)\kpc^{-1}\kms$. For the short-axis loop the mean frequencies are given by $\vOmega = (34.9,21.9,25.0)\kpc^{-1}\kms$ with errors $\Delta\vOmega=(0.6,0.1,0.2)\kpc^{-1}\kms$. For the long-axis loop the mean frequencies are given by $\vOmega = (36.9,22.4,24.0)\kpc^{-1}\kms$ with errors $\Delta\vOmega=(1.6,1.5,0.8)\kpc^{-1}\kms$.
We note that the frequency errors are largest at the turning points of the orbits for the loop orbits or near the centre of the potential for the box orbit.
\label{lastpage}

\bsp

\end{document}